\documentclass[preprint,prd,showpacs,amssymb,floatfix,12pt]{revtex4}
\usepackage{graphicx}
\usepackage[latin1]{inputenc}
\usepackage{amsmath}
\usepackage{amsfonts}
\usepackage{amssymb}
\usepackage{amsthm}
\usepackage{bm}
\usepackage{mathrsfs} 

\begin{document}
\newcommand{\scri}{\mathscr{I}}

\title{\bf Enhanced Black Hole Horizon Fluctuations}
\author{R. T. Thompson}
\email{robert@cosmos.phy.tufts.edu} 
\author{L.H. Ford}
\email{ford@cosmos.phy.tufts.edu}
\affiliation{Institute of Cosmology \\ 
Department of Physics and Astronomy \\
Tufts University, Medford, MA 02155}

\begin{abstract}
We discuss the possible
role of quantum horizon fluctuations on black hole radiance,
 especially whether they can invalidate Hawking's analysis based upon
transplanckian modes. We are particularly concerned with ``enhanced''
fluctuations produced by gravitons or matter fields in
squeezed vacuum states sent into the black hole after the collapse
process. This allows for the possibility of increasing
the fluctuations well above the vacuum level. We find that these
enhanced fluctuations could significantly alter stimulated emission,
but have little effect upon the spontaneous emission. Thus the thermal
character of the Hawking radiation is remarkably robust.
\end{abstract}


\maketitle

\baselineskip=13pt 

\section{Introduction}
Hawking's discovery~\cite{Hawking:1974sw} of black hole 
radiance has forged an elegant link
between relativity, quantum theory, and thermodynamics. However, some
unsolved problems remain, including the information and transplanckian
issues. The question of whether information is lost in the black hole
evaporation process has been vigorously debated by many authors. (See,
for example Ref.~\cite{Giddings:2006sj} and references therein.)
The transplanckian issue arises because
Hawking's original derivation used quantum field theory on a
fixed background spacetime and requires incoming vacuum modes with
frequencies far above the Planck scale. Alternative derivations have
been proposed, especially by Unruh~\cite{Unruh:1980cg,Unruh:1994je} 
and by Jacobson and 
coworkers~\cite{Jacobson:1991gr,Jacobson:1993hn,Jacobson:1996zs,
Corley:1996ar,Jacobson:2000gw} which involve cutoffs
and a non-linear dispersion relation. The non-linearity can lead the
the phenomenon of ``mode regeneration'', whereby the modes needed for
the outgoing particles are created just before they are needed, rather
than being redshifted transplanckian modes. However, this approach requires
a breakdown of local Lorentz invariance and hence new, as yet
unobserved, physics.

In this paper, we wish to consider the effects of quantum horizon
fluctuations on the Hawking process. This is a topic which has been
discussed by several authors from various viewpoints. An early discussion
was given by York~\cite{York:1983zb}, 
who used a semiclassical approach to estimate the
magnitude of the quantum metric fluctuations near the horizon. Ford and
Svaiter~\cite{Ford:1997zb} 
used York's estimate to treat fluctuations of the outgoing rays,
and concluded that Hawking's derivation does not seem to be altered by
vacuum fluctuations of linearized quantum gravity. 
Parentani~\cite{Parentani:2000ts} and 
Barabes~{\it et al}~\cite{Barrabes:2000fr} have discussed the
possibility that quantum fluctuations could
be the source of the non-linearity needed for the mode regeneration picture.

Most of the papers cited in the previous paragraph deal with active
fluctuations, the spacetime geometry fluctuations arising from quantization
of the dynamical degrees of freedom of gravity itself.
Another source of spacetime fluctuations are the quantum fluctuations of
matter field stress tensors, which cause passive fluctuations. There has
been an extensive discussion of both types of fluctuations in recent years
in various contexts, including both black hole
spacetimes~\cite{York:1983zb,Ford:1997zb,Parentani:2000ts,Barrabes:2000fr,
Hu:1999cc,Tuchin:1998fm,Wu:1999ea,Sorkin:1999yj,Hu:2007tq,Giddings:2007ie}
and more general spacetime geometry 
fluctuations~\cite{Phillips:2000jm,Borgman:2003dm,Ford:2000vm,Hu:2003qn,
Ford:2005sp,Thompson:2006qe}

In the present paper, we will examine some examples of both active and
passive fluctuations. However, our main interest will be in the possibility
of enhancing the geometry fluctuations above the vacuum level
by use of gravitons or matter fields in squeezed vacuum states. We consider
a Schwarzschild black hole formed by gravitational collapse, and
then suppose that wavepackets of gravitons or matter fields in squeezed
states are sent across the horizon. This will cause greater geometry
fluctuations than would occur in the vacuum states of these fields. The
key question which we wish to address is whether these fluctuations
significantly alter the outgoing modes which will carry the thermal
radiation to distant observers. The technique which we employ to study
the effects of geometry fluctuations is based upon the geodesic deviation
equation. This allows a gauge invariant treatment using the Riemann tensor
correlation function.

In Sect.~\ref{sec:Hawking}, 
we review Hawking's derivation of black hole evaporation, and discuss
the transplanckian issue. In Sect.~\ref{sec:Derivation}, we
develop some of the formalism
of fluctuations of a geodesic separation vector which will be used in
subsequent sections. We next turn to the case of active fluctuations. Before
considering gravitons, we first investigate a simplified model of ``scalar
gravitons'' in Sect.~\ref{sec:ScalarGraviton}. 
This model reproduced the essential physics of the
effects of gravitons, but with reduced complexity. The case of gravitons,
quantized linear perturbations of the Schwarzschild geometry, is treated in
Sect.~\ref{sec:Graviton}. We next turn to passive
fluctuation effects in Sect.~\ref{sec:ScalarField},
 where stress tensor fluctuations of a scalar
field  are treated. We give a unified analysis of the results of all three
models in Sect.~\ref{sec:NewDiscussion}, and offer our conclusions in 
Sect.~\ref{sec:NewConclusions}.

\section{Derivation of the Hawking Effect} \label{sec:Hawking}
\begin{figure}
 \centering
 \scalebox{.5}{\includegraphics{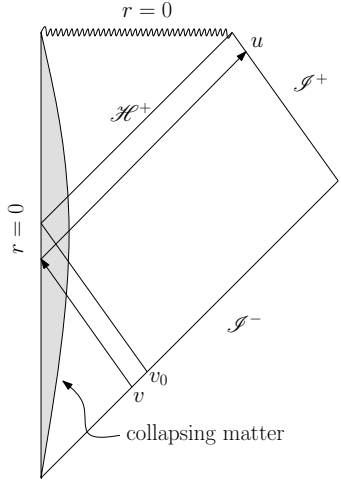}}\\
 \caption{The spacetime of a black hole formed by gravitational
    collapse is illustrated. The shaded region is the interior of the
collapsing body, the $r=0$ line on the left is the worldline of the center 
of this body, the $r=0$ line at the top of the diagram is the curvature
singularity, and $\mathscr{H}^+$ is the future event horizon.  An ingoing light
ray with $v < v_0$ from ${\scri^{-}}$ passes through the body and escapes
to ${\scri^{+}}$ as a $u = constant$ light ray. Ingoing rays with $v > v_0$
do not escape and eventually reach the singularity.}
 \label{Fig:BlackHoleST}
\end{figure}
In this section, we will briefly review Hawking's
derivation~\cite{Hawking:1974sw} of black hole
evaporation. The basic idea is to consider the spacetime of a black hole
formed by gravitational collapse, as illustrated in 
Figure~\ref{Fig:BlackHoleST}. 
 Here $v=t+r_*$, and $u=t-r_*$ are respectively the
ingoing and outgoing Eddington-Finkelstein coordinates, also referred to as the
advanced and retarded times, and $r_*=r+2M\ln (\tfrac{r}{2M}-1)$ is the tortoise
coordinate.
A quantum field
propagating in this spacetime is assumed to be in the in-vacuum state, that
is, containing no particles before the collapse.        In the case of a
massless field, a purely positive frequency mode proportional to
${\rm e}^{-i\omega v}$ leaves $\scri^{-}$, propagates through the
collapsing body, and reaches $\scri^{+}$ after undergoing a large redshift
in the region outside of the collapsing matter. At $\scri^{+}$,
the mode is now a mixture of positive and negative frequency parts,
signalling quantum particle creation. Of special interest are the modes
which leave $\scri^{-}$ just before the formation of the horizon, which is
the $v = v_0$ ray. These modes give the dominant contribution to the
outgoing flux at times long after the black hole has formed. After passing
through the collapsing body, they are $u = constant$ rays, where
\begin{equation}
u = -4M\, \ln\biggl(\frac{v_0 -v}{C}\biggr)\, , \label{eq:uofv}
\end{equation}
where $M$ is the black hole's mass, and $C$ is a constant.
The logarithmic dependence leads to a Planckian spectrum of created
particles. It also leads to the ``transplanckian issue'',  the enormous
frequency which the dominant modes must have when they leave $\scri^{-}$.
The typical frequency of the radiated particles reaching $\scri^{+}$
midway through the evaporation process is of order $1/M$, but the typical
frequency of these modes at $\scri^{-}$ is of order
\begin{equation}
\omega \approx M^{-1} {\rm e}^{(M/M_{Pl})^2}\, ,
\end{equation}
 where $M_{Pl}$ is the Planck mass.
Another way to state this is to note that the characteristic value of $u$
for these modes is of order
\begin{equation}
u_c \approx M \biggl(\frac{M}{m_p}\biggr)^2 \,.
\end{equation}
A geodesic observer who falls from rest at large distance from the black
hole will pass from $u=u_c$ to the horizon at $u = \infty$ in a proper time
of
\begin{equation}
\delta \tau \approx M\, e^{-{u_c}/{4M}} \approx
                    M\, e^{-{M^2}/{m_p^2}}\,.   \label{eq:del-tau}              
\end{equation}
which is far smaller than the Planck time. In this sense, the outgoing modes
are much less than a Planck length outside the horizon.

If spacetime geometry fluctuations cause these outgoing modes either
to be ejected prematurely, or to fall into the singularity, then the
outgoing radiation, and possibly the thermal character of the black
hole, could be greatly altered. This is the question which we wish to
address in this paper.

\section{Formalism} \label{sec:Derivation}
\subsection{Null Kruskal Coordinates}
Most of this work is done using null Kruskal coordinates, for which the line
element is
\begin{equation}
 ds^2 = -\frac{32M^3}{r} e^{-r/2M} dU dV + r^2 d\Omega^2
\end{equation}
where the coordinates $(U,V)$ are defined by
\begin{equation}
  U = -e^{-u/4M} \quad \mbox{and}\quad  V = e^{v/4M}
\end{equation}
and describe surfaces of constant phase, equivalently the path of light rays, in
a Schwarzschild space-time.   Kruskal coordinates are advantageous
because, unlike Schwarzschild
coordinates, they do not suffer a coordinate singularity at the horizon. 

In null Kruskal coordinates, $V$ is an affine parameter on $\mathscr{H}^+$ and
is approximately an affine parameter on outgoing null geodesics very near the
horizon, but only on that portion of the geodesic for which $UV \ll 1$, or
equivalently near the $r=2M$ surface.  However, outgoing null geodesics spend a
long affine time near $r=2M$ before finally escaping to infinity, so this should
be a good approximation for a large range of $V$.

Furthermore, on the past horizon of an eternal black hole, $\mathscr{H}^-$, the
Kruskal coordinate $U$ is an affine parameter for ingoing null
geodesics.  (See, for example, Eq.~12.5.10 in Ref.~\cite{Wald:1984rg}.)
Working in Kruskal coordinates and
using $U$ and $V$ as affine parameters near the past and future horizons,
$\ell^{\mu}=(0,1,0,0)$ and $s^{\mu}=(1,0,0,0)$ are tangent to, respectively,
outgoing and ingoing null geodesics near the horizon.

\subsection{Geodesic Deviation}

\begin{figure}
 \centering
 \scalebox{.5}{\includegraphics{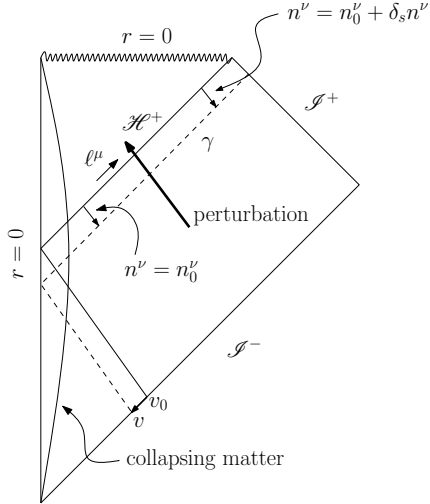}}\\
 \caption{Schwarzschild black hole space-time formed via gravitational collapse.
 The separation vector characterizing geodesic deviation of the horizon,
$\mathscr{H}^+$, and a nearby outgoing null geodesic, $\gamma$, is initially
$n^{\nu}=n^{\nu}_0$ but then evolves as $n^{\nu}=n^{\nu}_0 + \delta_s n^{\mu}$. 
The vector $\ell^{\mu}$ is tangent to the horizon. Perturbing fields
fall 
into the black hole well after the horizon forms.}
 \label{Fig:BlackHole2}
\end{figure}

The derivation of the Hawking effect outlined in
Sect.~\ref{sec:Hawking} requires propagation of a field from
$\scri^+$ backwards along a geodesic, through the collapsing body, and out to
$\scri^-$.  These tracked geodesics lie very close to the
horizon, and are separated from the horizon by some separation vector $n^{\mu}$
as in Fig. \ref{Fig:BlackHole2}.  In this figure, the geodesic (labeled
$\gamma$) appears to be a straight line at a fixed distance from the horizon so
that $n^{\mu}$ is constant.  This is not quite true; a particle following
$\gamma$ is eventually separated from the horizon by an infinite physical
distance.  Tracking the evolution of the separation vector from some initial
point out to $\scri^+$ requires integration of the geodesic deviation equation
from some initial point out to $\scri^+$.  One finds that the $U$ component of
$n^{\mu}$ is constant while the $V$ component is not.  Hawking is actually
considering only the fixed $U$ component as it is only $n^U$ that is relevant
for his derivation of Eq.~(\ref{eq:uofv}).

Consider $\mathscr{H}^+$, parametrized by $U=0$ and affine parameter
$\lambda_1=V$, and a nearby outgoing null geodesic just outside the horizon
parametrized by $U=n_0$ and affine parameter $\lambda_2\approx V$
for some small
$n_0 \ll 1$. Let $n^{\mu}$ connect points of equal affine parameter on
$\mathscr{H}^+$ and the geodesic with $U=n_0$.  Parameterizing the
geodesics such that initially $\lambda_1 = \lambda_2 = V_0$, the 
separation vector
is initially $n^{\mu}_0 = (n_0,0,0,0)$ at some point $V=V_0$.  In a flat
space-time the separation vector would not change and $n^{\mu}=n_0^{\mu}$
everywhere along the geodesic.   By applying the geodesic deviation
equation in Kruskal coordinates one can find the subsequent evolution of the
separation vector $n^{\mu}=n_0^{\mu}+\delta_sn^{\mu}$, where $\delta_s n^{\mu}$
represents the kinematic evolution of $n^{\mu}$ due to the classical
background.  If in addition, there is a perturbation of the background,
then $\bar n^{\mu}=n_0^{\mu}+\delta_sn^{\mu}+\delta_pn^{\mu}$, where
$\delta_pn^{\mu}$ represents the dynamical response of $\bar n^{\mu}$ to the
perturbation.  The bar on $\bar n^{\mu}$ is used to differentiate between the
background-only space-time separation vector and the background plus
perturbation space-time separation vector for this discussion.

\subsubsection{Background Space-Time}

Consider first the unperturbed Schwarzschild space-time of Fig.\
\ref{Fig:BlackHole2}.  Let $\ell^{\mu}=(0,1,0,0)$ be tangent to the horizon, and
let $n^{\mu}=n_0^{\mu}+\delta_sn^{\mu}$ denote the separation between the
horizon and a nearby outgoing null geodesic with initial separation $n^{\mu}_0$
at $V=V_0$ such that $n^{\mu}_0$ is null with $n^{\mu}\ell_{\mu}=1$.  It
suffices to choose $n^{\mu}_0=\left(n_0,0,0,0\right)$ with
$n_0=(g_{UV})^{-1}\vert_{r=2M} =1/(8 M^2)$.  
The evolution of the vector $n^{\mu}$
characterizes the geodesic deviation of these outgoing rays, and obeys the set
of differential equations
\begin{equation}
 \frac{D^2n^{\alpha}}{dV^2} =
R^{\alpha}_{\phantom{\alpha}\beta\mu\nu}\ell^{\beta}\ell^{\mu} n^{\nu}.
\end{equation}
Since $\ell^{\mu} = ({\partial}/{\partial V})^{\mu}$, the
geodesic deviation is
\begin{equation} \label{Eq:BackgroundDev}
\frac{D^2n^{\alpha}}{dV^2} = \frac{D^2(\delta_sn^{\alpha})}{dV^2} =
R^{\alpha}_{\phantom{\alpha}VV\nu}\ell^V\ell^V n^{\nu} \,.
\end{equation}
The only non-zero component of $R^{\alpha}_{\phantom{\alpha}VV\nu}$
is 
\begin{equation} \label{Eq:BackgroundRiemann1}
 R^V_{\phantom{V}VVU} =
 \frac{16M^3}{UVr^3}\left(1-\frac{2M}{r}\right)\, ,
\end{equation}
which near the horizon  reduces to
\begin{equation} \label{Eq:BackgroundRiemann2}
 R^V_{\phantom{V}VVU} \approx -2e^{-1}.
\end{equation}
Also near the horizon, the covariant derivative with respect to $V$ on the left
hand side of Eq.~(\ref{Eq:BackgroundDev}) reduces to an ordinary derivative,
which may be seen by direct calculation.  The second covariant derivative is
\begin{equation}
 \frac{D^2n^{\mu}}{dV^2} = \ell^{\gamma}\ell^{\sigma}
n^{\mu}_{\phantom{\mu},\sigma\gamma} +
\Gamma^{\mu}_{\delta\lambda,\gamma}\ell^{\gamma}\ell^{\delta} n^{\lambda} +
2\Gamma^{\mu}_{\delta\lambda}\ell^{\gamma}\ell^{\delta}
n^{\lambda}_{\phantom{\lambda},\gamma} +
\Gamma^{\mu}_{\gamma\beta}\Gamma^{\beta}_{\delta\lambda}\ell^{\gamma}\ell^{
\delta} n^{\lambda}.
\end{equation}
Each term with a Christoffel symbol is of the form $\Gamma^{\alpha}_{V\mu}$.  A
straightforward computation of the Christoffel symbols in null Kruskal
coordinates reveals that $\Gamma^{\alpha}_{V\mu}\to 0$ as $U\to 0$ for all
$\alpha$ and $\mu$.  The only non-trivial equation is then
\begin{equation}
 \frac{d^2(\delta_sn^{V})}{dV^2} = -2e^{-1}n_0
\end{equation}
which may be integrated from the initial point $V=V_0$, resulting in
\begin{equation} \label{Eq:BackgroundDeviation}
 n^{\mu} = n_0\left(1,-e^{-1}(V-V_0)^2\right) \quad \mbox{or} \quad
\left(n^{\mu}n_{\mu}\right)^2 = 4\left(g_{UV}\, n_0^2
e^{-1}\right)^2\left(V-V_0\right)^4 \,.
\end{equation}

\subsubsection{Perturbed Space-Time}
We wish to let the space-time fluctuate in some way and describe what happens to
the outgoing null geodesics.  Consider an ensemble of Schwarzschild space-times
such that the average apparent horizon is defined by $r=2M$ for some chosen
value of $M$.  Consider an outgoing geodesic just outside the average apparent
horizon.  For each space-time in the ensemble, define the separation vector
$\bar n^{\mu}$ such that in the average space-time $\bar n^{\mu} = n^{\mu}$
where $n^{\mu}$ is the classical solution for the Schwarzschild space-time of
mass $M$.

Let $\bar{g}_{\mu\nu}=g_{\mu\nu}+h_{\mu\nu}$ be the perturbed space-time, where
$g_{\mu\nu}$ is the unperturbed (Schwarzschild) background space-time and
$h_{\mu\nu}$ is the perturbation. The background metric is used to raise and
lower indices.  The Riemann tensor is calculated from
\begin{equation} \label{Eq:FullRiemann}
 \bar{R}^\alpha_{\phantom{\alpha}\beta\mu\nu} =
\bar{\Gamma}^\alpha_{\beta\nu,\mu}-\bar{\Gamma}^\alpha_{\beta\mu,\nu}+\bar{
\Gamma}^\alpha_{\mu\sigma}\bar{\Gamma}^{\sigma}_{\beta\nu}-
\bar{\Gamma}^\alpha_{\nu\sigma}\bar{\Gamma}^{\sigma}_{\beta\mu} \,,
\end{equation}
where
\begin{equation}
 \bar{\Gamma}^{\alpha}_{\mu\nu} = \frac{1}{2}
g^{\alpha\beta}\left(\bar{g}_{\beta\mu,\nu}+\bar{g}_{\beta\nu,\mu} -
\bar{g}_{\mu\nu,\beta}\right) 
\end{equation}
are the connection coefficients in the perturbed space-time. These may be
expanded to $\bar{\Gamma}^{\alpha}_{\mu\nu} = \Gamma^{\alpha}_{\mu\nu}
+\delta\Gamma^{\alpha}_{\mu\nu}$, where $\Gamma^{\alpha}_{\mu\nu}$ are the
connection coefficients of the background, and $\delta\Gamma^{\alpha}_{\mu\nu}$
is due to the perturbation.  This yields
\begin{equation} \label{Eq:TotalRiemann}
 \bar{R}^\alpha_{\phantom{\alpha}\beta\mu\nu} \approx
R^\alpha_{\phantom{\alpha}\beta\mu\nu}+\delta
R^\alpha_{\phantom{\alpha}\beta\mu\nu}
\end{equation}
to first order in the metric perturbation, where similarly
$R^\alpha_{\phantom{\alpha}\beta\mu\nu}$ and $\delta
R^\alpha_{\phantom{\alpha}\beta\mu\nu}$ denote the background and perturbation
contributions to the Riemann tensor. Let a semicolon denote covariant
differentiation with respect to the background One may then verify that
\begin{equation}\label{deltaR}
 \delta R^\alpha_{\phantom{\alpha}\beta\mu\nu} =
\delta\Gamma^\alpha_{\beta\nu;\mu}-\delta\Gamma^\alpha_{\beta\mu;\nu} \,,
\end{equation}
where
\begin{equation}\label{deltaGamma}
 \delta\Gamma^{\alpha}_{\mu\nu} = \frac{1}{2}
g^{\alpha\beta}\left(h_{\beta\mu;\nu}+h_{\beta\nu;\mu} -
h_{\mu\nu;\beta}\right).
\end{equation}

Let $\ell^{\mu}$ be fixed as tangent to the apparent horizon in the average (or
background) space-time. Let $\bar n^{\mu} =
n_0^{\mu}+\delta_sn^{\mu}+\delta_pn^{\mu}$ denote the separation vector, where
$n_0^\mu$ is the same initial separation as before, 
$\delta_s n^{\mu}$ is defined to
satisfy the background equation as above, and $\delta_p n^{\mu}$ encodes the
dynamical response of $\bar n^{\mu}$ to the perturbation. Letting $\bar D$ be
the covariant derivative with respect to the perturbed space-time, the geodesic
deviation is
\begin{multline} \label{Eq:PertGeoDev1}
\frac{\bar D^2\bar{n}^{\alpha}}{dV^2} = \frac{\bar
D^2(\delta_sn^{\alpha})}{dV^2} + \frac{\bar D^2(\delta_pn^{\alpha})}{dV^2} =
\bar R^{\alpha}_{\phantom{\alpha}\mu\beta\nu}\ell^{\mu}\ell^{\beta}\bar n^{\nu}
\\ = \left(R^{\alpha}_{\phantom{\alpha}\mu\beta\nu} + \delta
R^{\alpha}_{\phantom{\alpha}\mu\beta\nu}\right)\ell^{\mu}\ell^{\beta}\left(n_0^{
\nu} + \delta_sn^{\nu} + \delta_pn^{\nu}\right).
\end{multline}
Consider the left hand side of this equation.  The background terms involving
$\Gamma^{\alpha}_{V\mu}$ still vanish near the horizon.  There are, however,
terms involving $\delta\Gamma^{\alpha}_{V\mu}$ which do not vanish on the
horizon and which enter to first order in the metric perturbation.  In
particular, the first term on the left hand side is
\begin{multline}
 \frac{\bar D^2(\delta_sn^{\mu})}{dV^2} = \ell^{\gamma}\ell^{\sigma}
(\delta_sn^{\mu})_{,\sigma\gamma} +
\delta\Gamma^{\mu}_{\delta\lambda,\gamma}\ell^{\gamma}\ell^{\delta}
(\delta_sn^{\lambda}) +
2\delta\Gamma^{\mu}_{\delta\lambda}\ell^{\gamma}\ell^{\delta}
(\delta_sn^{\lambda})_{,\gamma} +
\delta\Gamma^{\mu}_{\gamma\beta}\delta\Gamma^{\beta}_{\delta\lambda}\ell^{\gamma
}\ell^{\delta}( \delta_sn^{\lambda}) \,,
\end{multline}
which contributes two terms to first order in the metric perturbation. 
Similarly for the second term on the left hand side:
\begin{multline}
 \frac{\bar D^2(\delta_pn^{\mu})}{dV^2} = \ell^{\gamma}\ell^{\sigma}
(\delta_pn^{\mu})_{,\sigma\gamma} +
\delta\Gamma^{\mu}_{\delta\lambda,\gamma}\ell^{\gamma}\ell^{\delta}
(\delta_pn^{\lambda}) +
2\delta\Gamma^{\mu}_{\delta\lambda}\ell^{\gamma}\ell^{\delta}
(\delta_pn^{\lambda})_{,\gamma} +
\delta\Gamma^{\mu}_{\gamma\beta}\delta\Gamma^{\beta}_{\delta\lambda}\ell^{\gamma
}\ell^{\delta} (\delta_pn^{\lambda}).
\end{multline}
However, $\delta_pn^{\lambda}$ is already first order in the metric
perturbation, so terms involving both $\delta_pn^{\lambda}$ and
$\delta\Gamma^{\alpha}_{\mu\nu}$ are second order.  Therefore to first order in
the metric perturbation, the left hand side of Eq.~(\ref{Eq:PertGeoDev1}) is
\begin{equation}
 \frac{\bar D^2\bar{n}^{\mu}}{dV^2} = \frac{d^2(\delta_sn)^{\mu}}{dV^2} +
\frac{d^2(\delta_pn^{\mu})}{dV^2} +
\delta\Gamma^{\mu}_{\delta\lambda,\gamma}\ell^{\gamma}\ell^{\delta}
(\delta_sn^{\lambda}) +
2\delta\Gamma^{\mu}_{\delta\lambda}\ell^{\gamma}\ell^{\delta}
(\delta_sn^{\lambda})_{,\gamma} \,,
\end{equation}
and the right hand side of Eq.~(\ref{Eq:PertGeoDev1}) is
\begin{equation}
 R^{\alpha}_{\phantom{\alpha}\mu\beta\nu}\ell^{\mu}\ell^{\beta}\left(n_0^{\nu} +
\delta_sn^{\nu}\right) +
R^{\alpha}_{\phantom{\alpha}\mu\beta\nu}\ell^{\mu}\ell^{\beta}(\delta_pn^{\nu})
+ \delta
R^{\alpha}_{\phantom{\alpha}\mu\beta\nu}\ell^{\mu}\ell^{\beta}\left(n_0^{\nu} +
\delta_sn^{\nu}\right).
\end{equation}
The first term is just the result obtained for the background.  Since
$\delta_sn^{\mu}$ is by definition the solution to 
\begin{equation}
 \frac{d^2(\delta_sn^{\alpha})}{dV^2} =
R^{\alpha}_{\phantom{\alpha}\mu\beta\nu}\ell^{\mu}\ell^{\beta}\left(n_0^{\nu} +
\delta_sn^{\nu}\right),
\end{equation}
these terms cancel.  Furthermore, since $\delta_sn^{\mu}$ has only a $V$
component and $\delta R^{\alpha}_{\phantom{\alpha}VVV}=0$ by the antisymmetry of
the Riemann tensor on the last two indices, then $\delta
R^{\alpha}_{\phantom{\alpha}\mu\beta\nu}\ell^{\mu}\ell^{\beta}(\delta_sn^{\nu})
= 0$ and Eq.\ (\ref{Eq:PertGeoDev1}) becomes
\begin{multline} \label{Eq:PertGeoDev2}
 \frac{d^2(\delta_pn^{\alpha})}{dV^2} =
R^{\alpha}_{\phantom{\alpha}\mu\beta\nu}\ell^{\mu}\ell^{\beta}(\delta_pn^{\nu})
+ \delta R^{\alpha}_{\phantom{\alpha}\mu\beta\nu}\ell^{\mu}\ell^{\beta}n_0^{\nu}
- \delta\Gamma^{\alpha}_{\delta\lambda,\gamma}\ell^{\gamma}\ell^{\delta}
(\delta_sn^{\lambda}) -
2\delta\Gamma^{\alpha}_{\delta\lambda}\ell^{\gamma}\ell^{\delta}
(\delta_sn^{\lambda})_{,\gamma}.
\end{multline}

The solution to this equation involves integrating twice over $V$ from some
initial point $V_0$.  It has already been found in 
Eq.~(\ref{Eq:BackgroundDeviation})
that $\delta_sn^{\lambda}\propto
(V-V_0)^2$.  We suppose that $\delta_pn^{\lambda}$ may also be expanded in
powers of $(V-V_0)$ and then approach the solution to Eq.\
(\ref{Eq:PertGeoDev2}) iteratively.  Suppose $\delta_pn^{\lambda} =
\delta_{p,1}n^{\lambda}+\delta_{p,2}n^{\lambda}$ in powers of $(V-V_0)$.  Since
the constant term is already accounted for in $n_0^{\lambda}$, then
$\delta_{p,1}n^{\lambda}$ must be $O(V-V_0)$ or smaller.  The double integral
over the $\delta_pn^{\lambda}$ and $\delta_sn^{\lambda}$ terms results in terms
of higher order in $(V-V_0)$, thus in the first iteration,
$\delta_{p,1}n^{\lambda}$ will be the solution to
\begin{equation}
 \frac{d^2(\delta_{p,1}n^{\alpha})}{dV^2} = \delta
R^{\alpha}_{\phantom{\alpha}\mu\beta\nu}\ell^{\mu}\ell^{\beta}n_0^{\nu},
\end{equation}
which turns out to be proportional to $(V-V_0)^2$.  The next iteration is the
solution to
\begin{equation}
 \frac{d^2(\delta_{p,2}n^{\alpha})}{dV^2} =
R^{\alpha}_{\phantom{\alpha}\mu\beta\nu}\ell^{\mu}\ell^{\beta}(\delta_{p,1}n^{
\nu}) - \delta\Gamma^{\alpha}_{\delta\lambda,\gamma}\ell^{\gamma}\ell^{\delta}
(\delta_sn^{\lambda}) -
2\delta\Gamma^{\alpha}_{\delta\lambda}\ell^{\gamma}\ell^{\delta}
(\delta_sn^{\lambda})_{,\gamma}.
\end{equation}
Each term on the right hand side is proportional to $(V-V_0)^2$, and the
solution is simply a double integral over $V$, giving a result
proportional to $(V-V_0)^4$.  Consequently, to lowest order in powers
of $(V-V_0)$, the solution to $\bar n^{\mu}$ for the perturbed space-time is
\begin{equation} \label{Eq:SeparationEvolution}
 \bar n^{\alpha} = n_0^{\alpha}+\delta_s n^{\alpha} + \int_{V_0}^{V}
dW\int_{V_0}^{W}dV\, \delta
R^{\alpha}_{\phantom{\alpha}\mu\beta\nu}\ell^{\mu}\ell^{\beta}n_0^{\nu}.
\end{equation}

To proceed further requires a model for fluctuations to be specified, and here
three different models will be considered in turn:
\begin{itemize}
\item A scalar graviton model where the metric is perturbed by a term
proportional to the product of a scalar field with the background metric, i.e.
$\bar{g}_{\mu\nu}=(1 + \Phi)g_{\mu\nu}$.
\item An ingoing gravitational wave actively perturbs the horizon in a
linearized theory of gravity. The perturbation to the Riemann tensor arises from
the perturbation to the metric.
\item An ingoing scalar field  provides a passive perturbation to the horizon. 
The perturbation to the Riemann tensor is the Ricci tensor contribution that
arises from the stress tensor of the scalar field.
\end{itemize}

In all three cases, the ingoing field is taken to occupy a squeezed quantum
state, $|\alpha, \zeta \rangle$.  In particular, the expectation value will be
evaluated with respect to a multimode squeezed vacuum state, $|0, \zeta \rangle
= \sum_{i=z_0}^{z_1}S(\zeta_i)|0\rangle$, which is further described in Appendix
\ref{app:SqueezedStates}. The excited modes will be taken to be
wavepackets which are sent into the black hole after the collapse, as
illustrated by the ingoing arrow in Fig.~\ref{Fig:BlackHole2}. 

It should be noted that the introduction of quantum fluctuations into
a black hole spacetime entails a significant conceptual
shift. Classical perturbations will shift the location of the horizon,
but do not change the fact that there is a precisely defined horizon.
Of course, the true event horizon in a classical spacetime, the light
ray which barely fails to escape to $\scri^+$, can only be
known when the complete history of the spacetime is known. Quantum
fluctuations introduce an additional ambiguity, whereby the precise
event horizon can never be known.

\subsubsection{Fluctuations}
Quantizing the ingoing perturbation field, $\delta_p n^{\mu}$ becomes a quantum
operator. To construct the operator $\delta_p\hat{n}^{\mu}$, consider
$\delta_p\hat{n}^{\mu}$ as the solution to
\begin{equation}
 \frac{d^2\delta_p \hat{n}^{\alpha}}{dV^2} = \delta
\hat{R}^{\alpha}_{\phantom{\alpha}\beta\mu\nu}\ell^{\beta}\ell^{\mu}n_0^{\nu}.
\end{equation}
Suppose we characterize fluctuations in the separation vector by the quantity
\begin{equation}
  \langle \bar{n}^{\mu}\bar{n}_{\mu}\rangle-\langle \bar{n}^{\mu}\rangle\langle
\bar{n}_{\mu}\rangle = \langle \delta_p \hat n^{\mu} \delta_p \hat
n_{\mu}\rangle - \langle \delta_p \hat n^{\mu} \rangle\langle \delta_p \hat
n_{\mu}\rangle.
\end{equation}
Due to the peculiarities of null Kruskal coordinates, this quantity is not a
good comparator for all three models.  In particular, it is identically zero for
the ``scalar graviton'' model, of order $(V-V_0)^4$ for the graviton model, and
of order $(V-V_0)^2$ for the scalar field model.  In order to compare the
results of the three models with each other and with the background result,
which is of order $(V-V_0)^2$, it is advantageous to consider instead the
variance
\begin{equation}
 \Delta(\bar n^{\mu}\bar n_{\mu})^2 =
\langle\left(\bar{n}^{\mu}(x)\bar{n}_{\mu}(x)\right)
\left(\bar{n}^{\mu}(x')\bar{n}_{\mu}(x')\right)\rangle-\langle\left(\bar{n}^{\mu
}(x)\bar{n}_{\mu}(x)\right)\rangle\langle
\left(\bar{n}^{\mu}(x')\bar{n}_{\mu}(x')\right)\rangle.
\end{equation}
It is straightforward to show that
\begin{multline}
 \Delta(\bar n^{\mu}\bar n_{\mu})^2 =
4\left(n^{\mu}_0(x)n^{\nu}_0(x')+n^{\mu}_0(x)\delta_s n^{\nu}(x')+\delta_s
n^{\mu}(x)\delta_s n^{\nu}(x')\right)\\ \times \left[\langle\delta_p \hat
n_{\mu}(x) \delta_p \hat n_{\nu}(x')\rangle - \langle\delta_p \hat
n_{\mu}(x)\rangle \langle \delta_p \hat n_{\nu}(x')\rangle \right] +
2\left(n^{\mu}_0(x)+\delta_s n^{\mu}(x)\right) \\ \times \left[\langle \delta_p
\hat n_{\mu}(x)\delta_p \hat n^{\nu}(x')\delta_p \hat n_{\nu}(x')\rangle -
\langle \delta_p \hat n_{\mu}(x)\rangle\langle\delta_p \hat n^{\nu}(x')\delta_p
\hat n_{\nu}(x')\rangle \right] \\ + \langle \delta_p \hat n^{\mu}(x)\delta_p
\hat n_{\mu}(x)\delta_p \hat n^{\nu}(x')\delta_p \hat n_{\nu}(x')\rangle  -
\langle \delta_p \hat n^{\mu}(x)\delta_p \hat n_{\mu}(x)\rangle\langle\delta_p
\hat n^{\nu}(x')\delta_p \hat n_{\nu}(x')\rangle.
\end{multline}
As will be demonstrated, each $\delta_p \hat n_{\nu}(x)\propto (V-V_0)^2$; 
additionally, $\delta_s n_{\nu}(x)\propto (V-V_0)^2$.  Therefore, to lowest
order in $(V-V_0)$ we have
\begin{multline} \label{Eq:variance}
 \Delta(\bar n^{\mu}\bar n_{\mu})^2 = 4\left[\langle \left(n^{\mu}_0(x)\delta_p
\hat n_{\mu}(x)\right)\left(n^{\nu}_0(x')\delta_p \hat n_{\nu}(x')\right)\rangle
\right. \left. - \langle n^{\mu}_0(x)\delta_p \hat n_{\mu}(x)\rangle\langle
n^{\nu}_0(x')\delta_p \hat n_{\nu}(x')\rangle\right].
\end{multline}
This then is the primary quantity of interest to calculate for the three
fluctuation models.


\section{Scalar Graviton Model} \label{sec:ScalarGraviton}
Turn now to the scalar graviton model, where the perturbation is simply a scalar
field, $\Phi$, (a dilaton) multiplying the background metric, 
 $\bar g_{\mu\nu} = \left( 1+\Phi\right)g_{\mu\nu}$.  Here
$g_{\mu\nu}$ is the unperturbed (Schwarzschild) background space-time metric and
we may define $\bar g^{\mu\nu}=g^{\mu\nu}(1-\Phi)$ such that to first order in
$\Phi$, $\bar g^{\mu\alpha}\bar
g_{\mu\beta}=\delta^{\alpha}_{\phantom{\alpha}\beta}$. The scalar field $\Phi$
will be a free quantum scalar field (multiplied by $\ell_P$) and
will be defined in the following section.  One may object that this model is
simply a conformal transformation of the space-time under which the light cone
structure remains invariant.  However, while it is true that the light cone is
invariant under a conformal transformation, the geodesic deviation is 
affected.  This is because the Riemann tensor involves derivatives of the
conformal factor, so that the Riemann tensor of the transformed space-time is
not equal to a simple conformal transformation of the Riemann tensor.  For
example, Robertson-Walker space-time is conformally flat but has non-trivial
geodesic deviation.  This model is useful as a simplified model which
reproduces the essential features of the more complicated graviton
model of Sect.~\ref{sec:Graviton}.

\subsection{Normalized Wave Packets}
\label{sec:normalized}
Let a scalar field propagate from $r_* = \infty$, through the potential barrier
of the black hole, to the horizon at $r_*=-\infty$.  The wave function must
satisfy the wave equation in the Schwarzschild geometry.  Following Hawking
\cite{Hawking:1974sw}, Fourier decompose solutions of the wave equation with
respect to advanced or retarded time, use continuum normalization, and expand in
spherical harmonics.  Using ingoing Eddington-Finkelstein coordinates, a single
ingoing mode is
\begin{equation}
 \psi_{\omega \ell m} = \frac{Y_{\ell m}(\theta,\varphi)}{r\sqrt{2\pi\omega}}
F_{\omega}(r) e^{-i\omega v}
\end{equation}
from which ingoing wave packets may be constructed as
\begin{equation} \label{Eq:WavePacket}
 \psi_{j n} = \varepsilon_j^{-\frac{1}{2}}
\int_{j\varepsilon_j}^{(j+1)\varepsilon_j} e^{-2\pi i n \omega/\varepsilon_j}
\psi_{\omega \ell m} \, d\omega.
\end{equation}
The integer $j$ controls where in frequency space the wave packet is peaked,
while $\varepsilon_j$ controls the width of the wavepacket and has units of
frequency.  The integer $n$ describes which wave packet is under consideration. 
This construction allows for wave packets to be sent in at regular intervals of
${2\pi}/{\varepsilon_j}$ with various frequencies.  Thus $\psi_{jn}$ is the
$n^{th}$ wave packet sent in with component frequencies ranging from
$j\varepsilon_j$ to $(j+1)\varepsilon_j$.  The function $F_{\omega}(r)$ is in
general a complex function which depends in some complicated way on the geometry
of the space-time.  For sharply peaked wave packets, $F_{\omega}(r)$ is of order
unity at infinity, and essentially reduces to a transmission coefficient near
the horizon.  In Appendix~\ref{app:ScalarNormalization}, it is shown that these
wavepackets are properly normalized.
The quantized scalar field is constructed from the wavepackets defined above as
\begin{equation}
 \Phi = \sum_{jn} \left(\psi_{jn} \hat{a}_{jn} + \psi_{jn}^*
\hat{a}_{jn}^{\dagger}\right).
\end{equation}
In general it suffices to consider a single ingoing wavepacket, thus in what
follows the index $n$ will usually be suppressed and assumed fixed.

\subsection{Fluctuations}
Since the metric of the full space-time obeys the same symmetries as the
background space-time, it is straightforward to calculate the Riemann tensor
exactly from Eq.\ (\ref{Eq:FullRiemann}). In particular, to first order in
$\Phi$ and its derivatives, one finds the relevant quantity 
[see Eq.~(\ref{Eq:TotalRiemann})]
\begin{equation}
\delta R^V_{\phantom{V}VVU} = \Phi_{,VU}
\end{equation}
Applying Eq.~(\ref{Eq:SeparationEvolution}) one finds, 
to leading order in $(V-V_0)$,
\begin{equation}
 n_0^{\mu}(x)\delta_p n_{\mu}(x) = 
g_{UV}\, n_0^2 \int_{V_0}^V dW \int_{V_0}^W dV
\, \delta R^{V}_{\phantom{V}VVU} = 
g_{UV}\, n_0^2 \int_{V_0}^V dW \int_{V_0}^W dV
\, \Phi_{,VU}.
\end{equation}
Recall that $g_{UV} =8 M^2$ near $r=2M$.
Clearly, $\langle \Phi\rangle =0$ even when evaluating the expectation value
with respect to a squeezed vacuum state; therefore the variance, Eq.\
(\ref{Eq:variance}), becomes
\begin{equation}
 \Delta(\bar n^{\mu}\bar n_{\mu})^2 = 4\langle \left(n^{\mu}_0(x)\delta_p 
n_{\mu}(x)\right)\left(n^{\nu}_0(x')\delta_p  n_{\nu}(x')\right)\rangle.
\end{equation}
Expanding in terms of the mode functions, one finds
\begin{multline}
 \Delta(\bar n^{\mu}\bar n_{\mu})^2 = 4(g_{UV}\,n^2_0)^2\sum_{j,k}\int_{V_0}^V
dW \int_{V_0}^W dV \int_{V_0}^V dW' \int_{V_0}^{W'} dV' \\ \times \left\{
\psi_{j,VU}(x)\, \psi_{k,V'U'}(x')\, \langle\zeta,0|\hat a_{j}\hat
a_{k}|0,\zeta\rangle  +  \psi_{j,VU}(x)\, \psi^*_{k,V'U'}(x')\,
\langle\zeta,0|\hat a_{j}\hat a^{\dagger}_{k}|0,\zeta\rangle \right. \\ \left. +
\psi^*_{j,VU}(x)\, \psi_{k,V'U'}(x')\, \langle\zeta,0|\hat a^{\dagger}_{j}\hat
a_{k}|0,\zeta\rangle + \psi^*_{j,VU}(x)\, \psi^*_{k,V'U'}(x')\,
\langle\zeta,0|\hat a^{\dagger}_{j}\hat a^{\dagger}_{k}|0,\zeta\rangle \right\}
\end{multline}
where we choose to evaluate the expectation value with respect to a multimode
squeezed vacuum state $|0,\zeta\rangle = \prod_{i=z_0}^{z_1}
S(\zeta_i)|0\rangle$.  Using the results of Appendix \ref{app:SqueezedStates},
this becomes
\begin{multline}
 \Delta(\bar n^{\mu}\bar n_{\mu})^2 = 4(g_{UV}\,n^2_0)^2\sum_{j,k} \int_{V_0}^V
dW \int_{V_0}^W dV \int_{V_0}^V dW' \int_{V_0}^{W'} dV' \\ \times
\delta_{jk}\left\{
\psi_{j,VU}(x)\, \psi_{k,V'U'}(x')\,
\left(1+\Theta_{z_0z_1}(j)(\cosh\rho_j-1\right)\left(-\Theta_{z_0z_1}
(k)\sinh\rho_k\right) \right. \\
+  \psi_{j,VU}(x)\, \psi^*_{k,V'U'}(x')\,
\left(1+\Theta_{z_0z_1}(j)(\cosh\rho_j-1\right)
\left(1+\Theta_{z_0z_1}(k)(\cosh\rho_k-1\right) \\
+ \psi^*_{j,VU}(x)\, \psi_{k,V'U'}(x')\,
\left(-\Theta_{z_0z_1}(j)\sinh\rho_j\right)\left(-\Theta_{z_0z_1}
(k)\sinh\rho_k\right) \\ \left.
+ \psi^*_{j,VU}(x)\, \psi^*_{k,V'U'}(x')\,
\left(-\Theta_{z_0z_1}(j)\sinh\rho_j\right)\left(1+\Theta_{z_0z_1}
(k)(\cosh\rho_k-1\right) \right\} \, ,
\end{multline}
with the integer step function
\begin{equation}
  \Theta_{z_0 z_1}(j) = 
 \begin{cases}
 1, & z_0 \leq j \leq z_1, \\ 
 0, & \text{otherwise}.
 \end{cases}
\end{equation}

Notice that the factor multiplying $\psi_{j,VU}(x)\, \psi^*_{k,V'U'}(x')$
contains a $\delta_{jk}$ which makes $\Delta(\bar n^{\mu}\bar n_{\mu})^2$
divergent.  We therefore take renormalization to correspond to restricting the
sum over modes to those occupying an excited squeezed state mode.  For the
current model and the graviton model, this restriction corresponds to normal
ordering. In general, there will also be a vacuum contribution which is
being ignored here. This should be a good approximation for highly
excited states, that is, 
the limit of large squeeze parameter, $\cosh\rho_j\approx\sinh\rho_j\approx
{e^{\rho_j}}/{2}$.  Taking this limit and
restricting the sum to $z_0 \leq j,k \leq z_1$, the integer step function
$\Theta_{z_0z_1} =1$ and this simplifies to
\begin{equation}
 \Delta(\bar n^{\mu}\bar n_{\mu})^2 = (g_{UV}\,n^2_0)^2\sum_{j=z_0}^{z_1}
\left|e^{\rho_j} \int_{V_0}^V dW \int_{V_0}^W dV \left(\psi_{j,VU}(x) -
\psi_{j,VU}^*(x)\right)\right|^2.
\end{equation}
The $V$ integral is trivial and we have
\begin{equation} \label{Eq:SGVariance1}
 \Delta(\bar n^{\mu}\bar n_{\mu})^2 = (g_{UV}\,n^2_0)^2\sum_{j=z_0}^{z_1}
\left|e^{\rho_j} \int_{V_0}^V dW \left(\psi_{j,U}\big|_{V=V_0}^W -
\psi_{j,U}^*\big|_{V=V_0}^W\right)\right|^2.
\end{equation}
Consider $\psi_{j,U}(x)$.  Near the horizon, $F_{\omega}(r)$ reduces to a
transmission coefficient which depends only on $\omega$ and $\ell$.  One may
then find
\begin{equation}
 \psi_{j,U}\approx \int_{j\varepsilon_j}^{(j+1)\varepsilon_j} d\omega \,
\frac{Y_{\ell m} e^{-1} e^{-i \omega\delta_j}}{2M\sqrt{2\pi\varepsilon_j\omega}}
F(\omega) V^{1-4i M\omega}\,,
\end{equation}
where $\delta_j=2\pi n/\varepsilon_j$. We recall that
$(1-\tfrac{2M}{r})\approx -UVe^{-1}$, and use the definition of $V$ to write
$Ve^{-i\omega v} =V^{1-4iM\omega}$.  Next, one finds
\begin{multline} \label{Eq:PreExpansion}
 \int_{V_0}^V dW \psi_{j,U}(x)\big|_{V=V_0}^{V=W} = 
\int_{j\varepsilon_j}^{(j+1)\varepsilon_j} d\omega\, \frac{e^{-1}Y_{\ell
m}F(\omega)e^{-i\omega\delta_j}}{2M\sqrt{2\pi\omega\varepsilon_j}} \\ \times
\left[(2-4iM\omega)^{-1}\left(V^{2-4iM\omega}-V_0^{2-4iM\omega}\right)-V_0^{
1-4iM\omega}(V-V_0)\right].
\end{multline}
Expanding the bracketed terms in powers of $(V-V_0)$, this becomes
\begin{equation}
 \int_{V_0}^V dW \psi_{j,U}(x)\big|_{V=V_0}^{V=W} =
\int_{j\varepsilon_j}^{(j+1)\varepsilon_j} d\omega \, \frac{e^{-1}Y_{\ell
m}F(\omega)(1-4iM\omega)}{4M\sqrt{2\pi\omega\varepsilon_j}}
e^{-i\omega(v_0+\delta_j)}(V-V_0)^2.  \label{eq:packetint}
\end{equation}
Here $v_0$ is the Eddington-Finkelstein coordinate corresponding to $V_0$, i.e.\
$V_0 = e^{v_0/4M}$.  Use Eq.~(\ref{eq:packetint}) 
in Eq.\ (\ref{Eq:SGVariance1}) and let $m=0$,
then $ Y^*_{\ell 0} = Y_{\ell 0}$ and the spherical harmonics factor out. Using
this and the results for the classical deviation, Eq.\
(\ref{Eq:BackgroundDeviation}), and inserting the appropriate powers of the
Planck length, $\ell_P$, the fractional fluctuations are found to be
\begin{multline} \label{Eq:ScalarGravFluc}
 \frac{\Delta(\bar n^{\mu}\bar n_{\mu})^2}{(n^{\mu}n_{\mu})^2} =
\sum_{j=z_0}^{z_1} \Bigg|\int_{j\varepsilon_j}^{(j+1)\varepsilon_j} d\omega \,
\frac{e^{\rho_j} Y_{\ell 0}\ell_P }{8M\sqrt{2\pi\omega\varepsilon_j}} \\ \times
\left[\left(F(\omega)e^{-i\omega(v_0+\delta_j)} -
F^*(\omega)e^{i\omega(v_0+\delta_j)}\right) -
4iM\omega\left(F(\omega)e^{-i\omega(v_0+\delta_j)} +
F^*(\omega)e^{i\omega(v_0+\delta_j)}\right)\right]\Bigg|^2.
\end{multline}

Technically, expanding Eq.\ (\ref{Eq:PreExpansion}) in powers of $(V-V_0)$ is
really an expansion in powers of $2M\omega(V-V_0)$.  In considering the result
for the fractional fluctuations, therefore, one must bear in mind that we are
considering the limit where $(V-V_0)<1$ and also $\omega \leq (2M)^{-1}$.  In
the limit $\omega\to0$, $F(\omega)\to B_{\ell}(2i\omega M)^{\ell+1}$ where
$\vert B_{\ell}\vert$ is of order 1.  In this case one finds
\begin{equation} \label{Eq:SGLowFreq}
 \frac{\Delta(\bar n^{\mu}\bar n_{\mu})^2}{(n^{\mu}n_{\mu})^2} = \frac{\vert
B_{\ell}\vert^2 Y^2_{\ell 0}}{8}\sum_{j=z_0}^{z_1} \frac{e^{2\rho_j} \ell^2_P
(2M)^{2\ell}}{\pi\varepsilon_j} \left[\int_{j\varepsilon_j}^{(j+1)\varepsilon_j}
d\omega \, \omega^{\ell+1/2} \cos\omega(v_0+\delta_j)\right]^2.
\end{equation}
Near $\omega \approx (2M)^{-1}$, the transmission coefficient is of order 1. It
follows that if the wavepacket is sharply peaked near $\omega \approx (2M)^{-1}$
\begin{equation} \label{Eq:SGHighFreq}
 \frac{\Delta(\bar n^{\mu}\bar n_{\mu})^2}{(n^{\mu}n_{\mu})^2} = \frac{Y^2_{\ell
0}}{16} \sum_{j=z_0}^{z_1} \frac{e^{2\rho_j}\ell^2_P}{M\pi\varepsilon_j}
\left[\int_{j\varepsilon_j}^{(j+1)\varepsilon_j} d\omega \,
\sin\omega(v_0+\delta_j)+ 2\cos\omega(v_0+\delta_j)\right]^2.
\end{equation}
Let us leave further analysis until Section \ref{sec:NewConclusions} and
first consider the other models.


\section{Graviton Model}\label{sec:Graviton}
The next model for fluctuations is that of an ingoing gravitational wave
occupying a squeezed state.  In this case the perturbation field mode functions
are constructed from the allowed classical black hole perturbations.

\subsection{Even Parity Classical Perturbations}
With Regge and Wheeler paving the way, the subject of classical black hole
perturbations was thoroughly studied by Vishveshwara, Eddlestein, Zerilli,
Price, Tuekolsky, and others
\cite{Regge:1957td,PhysRevD.1.3514,PhysRevD.1.2870,Zerilli:1970se,
PhysRevD.2.2141,PhysRevD.5.2419,PhysRevD.5.2439,PhysRevLett.29.1114}.  We begin
with the original formulation of metric perturbations by Regge and Wheeler. 
These come in two varieties, even and odd parity.  In this work, we
will give an explicit treatment for the even parity case. However,
gravitons in odd parity wavepackets can be shown to lead to similar
conclusions as we will find here.  Purely even parity waves are physically
realizable, being generated, for example, by matter falling radially into a
black hole \cite{PhysRevD.2.2141}.  In Schwarzschild coordinates and using the
Regge-Wheeler gauge, the even parity metric perturbation is
\begin{equation} \label{Eq:EvenParity}
  \Psi_{\mu\nu} = e^{-i\omega t} P_{\ell}(\cos\theta) \left(
  \begin{smallmatrix}
   (1-\tfrac{2M}{r})H_0(r) & H_1(r) & 0 & 0\\
   H_1(r) & (1-\tfrac{2M}{r})^{-1}H_2(r) & 0 & 0\\
   0 & 0 & r^2K(r) & 0\\
   0 & 0 & 0 & r^2\sin^2\theta K(r)
  \end{smallmatrix} \right)
\end{equation}
While this equation pertains to a particular choice of gauge, our results are
based on the Riemann tensor correlation function and as such are gauge
invariant.  Zerilli \cite{Zerilli:1970se, PhysRevD.2.2141} showed that the even
parity radial functions $H_0(r),H_1(r),H_2(r)$, and $K(r)$ may be related to a
new radial function, $Z(r)$, that obeys a single Schr\"odinger-type equation
\begin{equation} \label{Eqn:ZerilliEqn}
 \frac{d^2 Z(r)}{dr_*^2} + (\omega^2 - V_{eff})Z(r) = 0,
\end{equation}
with an effective potential
\begin{equation} \label{Eqn:ZerilliPotential}
 V_{eff} = \left(\frac{1-2M}{r}\right)\frac{2\lambda^2(\lambda+1)r^3 +
6\lambda^2 M r^2 + 18\lambda M^2 r + 18M^3}{r^3(\lambda r + 3M)^2}\,,
\end{equation}
where $\lambda= (\ell-1)(\ell+2)/2$.

Equation (\ref{Eq:EvenParity}) may be transformed to null Kruskal coordinates,
which we indicate with a superscript, $\Psi_{\mu\nu}^{(K)}$.  Next, construct
the wavepackets
\begin{equation}
 \Psi_{(jn)\mu\nu}^{(K)}=\int_{j\varepsilon_j}^{(j+1)\varepsilon_j}d\omega\,
A_{\ell m}(\omega)e^{-i\omega\delta_j}\Psi_{\mu\nu}^{(K)} \, ,
\end{equation}
where $\delta_j = 2\pi n/\varepsilon_j$ and $A_{\ell m}(\omega)$ is a
normalization factor.  The normalization of this perturbation mode requires some
amount of work, which is relegated to Appendix \ref{app:GravNormalization}. The
ingoing linearly quantized graviton perturbation field is then
\begin{equation} \label{Eq:GravPertField}
 h^{(K)}_{\mu\nu}=\sum_{jn}\left(\Psi_{(jn)\mu\nu}^{(K)}\hat{a}_{jn}
+H.C.\right).
\end{equation}
The interpretation of the integers $j$ and $n$ is exactly the same as 
discussed in Sect.~\ref{sec:normalized}.  Again, it suffices
to consider $n$ fixed, so the index $n$ will in general be suppressed. Using
the results of Appendix \ref{app:GravNormalization}, the properly normalized
even parity wavepacket is written
\begin{equation}
  \Psi_{(j)\mu\nu} = \int_{j\varepsilon_j}^{(j+1)\varepsilon_j}
\frac{d\omega}{\sqrt{\pi\tilde{L}\varepsilon_j\omega}} e^{-i\omega\delta_j}
\Psi_{\mu\nu} \, ,
\end{equation}
where
\begin{equation}
 \tilde{L}=\frac{1}{2\ell+1}\left[2+\ell^2(\ell+1)(\ell+3)\right] -
\frac{(\ell+1)!}{(\ell-1)!}.
\end{equation}
With this normalization, it is understood that the Zerilli radial function takes
the asymptotic value $Z(r_*\to\infty) = e^{-i\omega r_*}$

\subsection{Fluctuations}
{} From Eqs.\ (\ref{deltaGamma}) and (\ref{Eq:GravPertField}), one finds
\begin{equation}
 \delta\hat\Gamma^{\alpha}_{\mu\nu} =
\frac{1}{2}g^{\alpha\beta}\sum_{j}\int_{j\varepsilon_j}^{(j+1)\varepsilon_j}
d\omega\, \left[A_{\ell
m}(\omega)e^{-i\omega\delta_j}\left(\Psi_{\mu\beta;\nu}+\Psi_{\beta;\nu\mu}
-\Psi_{\mu\nu;\beta}\right)\hat{a}_{j}+H.C.\right]\,,
\end{equation}
and by extension
\begin{equation}
 \delta \hat R^{\alpha}_{\phantom{\alpha}\beta\mu\nu} =
\sum_{j}\int_{j\varepsilon_j}^{(j+1)\varepsilon_j}d\omega\, \left[A_{\ell
m}(\omega)e^{-i \omega\delta_j} \delta
R^{\alpha}_{\phantom{\alpha}\beta\mu\nu}\hat{a}_{j}+H.C.\right].
\end{equation}
Here the hat has been reinserted to clarify that the operators
$\delta\hat\Gamma^{\alpha}_{\mu\nu}$ and $\delta \hat
R^{\alpha}_{\phantom{\alpha}\beta\mu\nu}$ may be constructed from the classical
quantity corresponding to a classical single mode perturbation. Consequently, 
\begin{equation}
 \delta_p\hat{n}^{\mu} =
\sum_{j}\int_{j\varepsilon_j}^{(j+1)\varepsilon_j}d\omega\, \left[A_{\ell
m}(\omega)e^{-i\omega\delta_j}\delta_p n^{\mu} \hat{a}_{j}+H.C.\right].
\end{equation}

The second term of Eq.\ (\ref{Eq:variance}) is zero for squeezed
vacuum states. 
Recognizing that $A_{\ell m}(\omega)$ is real, the variance for the graviton
model is
\begin{multline}
 \Delta(\bar n^{\mu} \bar n_{\mu})^2 =
4\langle\left(n^{\mu}_0(x)\delta_p\hat{n}_{\mu}(x)\right)
\left(n^{\nu}_0(x')\delta_p\hat{n}_{\nu}(x')\right)\rangle = \\
4\sum_{j}\sum_{k}\int_{j\varepsilon_j}^{(j+1)\varepsilon_j}d\omega
\int_{k\varepsilon_k}^{(k+1)\varepsilon_k}d\omega' A_{\ell m}(\omega)A_{\ell
m}(\omega') \\
 \left\{e^{-i \left(\omega\delta_j+\omega'\delta_k\right)}
\left(n^{\mu}_0(x)\delta_pn_{\mu}(x)\right)\left(n^{\nu}_0(x')\delta_pn_{\nu}
(x')\right)\langle\hat{a}_{j}\hat{a}_{k}\rangle \right.
 \\ + e^{i \left(\omega\delta_j+\omega'\delta_k\right)}
\left(n^{\mu}_0(x)\delta_pn_{\mu}^*(x)\right)\left(n^{\nu}_0(x')\delta_pn^*_{\nu
}(x')\right)\langle\hat{a}_{j}^{\dagger}\hat{a}_{k}^{\dagger}\rangle
 \\+ e^{-i \left(\omega\delta_j-\omega'\delta_k\right)}
\left(n^{\mu}_0(x)\delta_pn_{\mu}(x)\right)\left(n^{\nu}_0(x')\delta_pn^*_{\nu}
(x')\right)\langle\hat{a}_{j}\hat{a}_{k}^{\dagger}\rangle
 \\ +\left. e^{i \left(\omega\delta_j-\omega'\delta_k\right)}
\left(n^{\mu}_0(x)\delta_pn^*_{\mu}(x)\right)
\left(n^{\nu}_0(x')\delta_pn_{\nu}(x')\right)
\langle\hat{a}_{j}^{\dagger}\hat{a}_{k}\rangle \right\}.
\end{multline}
We choose to evaluate the expectation value with respect to a multimode squeezed
vacuum state $|0,\zeta\rangle = \prod_{i=z_0}^{z_1} S(\zeta_i)|0\rangle$.  As
discussed for the scalar graviton, renormalization
amounts to restricting the sum to those states which lie in the range of
squeezing.  Together with the results of Appendix \ref{app:SqueezedStates} and
in the limit of large squeeze parameter, $\rho$, this gives
 \begin{equation} \label{Eq:expectation}
 \Delta(\bar n^{\mu} \bar n_{\mu})^2 =  \sum_{j=z_0}^{z_1} e^{2\rho_j}
\left|\int_{j\varepsilon_j}^{(j+1)\varepsilon_j}d\omega A_{\ell m}(\omega)
\left(e^{-i\omega\delta_j} n^{\mu}_0(x)\delta_pn_{\mu}(x)-
e^{i\omega\delta_j} n^{\mu}_0(x)\delta_pn_{\mu}^*(x)\right)\right|^2.
\end{equation}

It remains to calculate $n^{\mu}_0\delta_pn_{\mu}$ due to a single mode
classical perturbation.  Unfortunately this is difficult to do analytically, but
is possible with the use of a computer algebra program.  To further simplify the
problem we restrict attention to perturbations of purely even parity with
$\ell=2$.
To calculate $\delta_pn^{\mu}$ near the horizon, one must first calculate
$\delta R^{\alpha}_{\phantom{\alpha}\beta\mu\nu}$ near the horizon in null
Kruskal coordinates.  There are several possible routes toward obtaining this
information.  The most straightforward might appear to be to first transform the
metric perturbation to Kruskal coordinates and then proceed from 
Eq.~(\ref{deltaR}).  
The difficulty with this method arises when one tries to take
the limit of $\delta R^{\alpha}_{\phantom{\alpha}\beta\mu\nu}$ near the horizon.
 In null Kruskal coordinates, the approach to the horizon is along a line of
constant $V$ rather than a line of constant $t$.  The equations, however, still
involve $r$, implicitly defined in terms of $U$ and $V$, so taking the limit
is not a well defined operation.

It is instead simpler to begin with the metric perturbation in Schwarzschild
coordinates and compute $\delta R^{\alpha}_{\phantom{\alpha}\beta\mu\nu}$ via
Eqs.\ (\ref{deltaR}) and (\ref{deltaGamma}) for a classical single mode ingoing
perturbation.  Expressing the radial functions $H(r)$, $H_1(r)$, and $K(r)$ in
terms of the Zerilli function, $Z(r)$, one then uses $Z(r)$ to first order in
$(r-2M)$ to expand $\delta R^{\alpha}_{\phantom{\alpha}\beta\mu\nu}$ in powers
of $(r-2M)$ near the horizon. It turns out to be necessary to expand the Zerilli
function to at least $O(r-2M)$ to ensure that the metric perturbation remains
finite on the horizon.  Since the initial calculation is done in Schwarzschild
coordinates, the limit is well defined as $r\to 2M$ along a line of constant
$t$.  Once $\delta R^{\alpha}_{\phantom{\alpha}\beta\mu\nu}$ is known near the
horizon, it may then be transformed to null Kruskal coordinates.

Using this procedure to solve for $\delta_pn^{\mu}$ as expressed in 
Eq.\ (\ref{Eq:SeparationEvolution}), one may go on to find
\begin{multline}
  n^{\mu}_0(x)\delta_pn_{\mu}(x) = 
-(g_{UV}e^{-1}n_0^2)(V-V_0)^2\frac{T(\omega)e^{-i\omega
v_0}}{324M(i+4M\omega)}\left(11800i \right. \\ \left. + 34321M\omega 
-39783iM^2\omega^2+14348M^3\omega^3\right)\left(1+3\cos2\theta\right).
\end{multline}
Inserting the appropriate powers of the Planck length, the fractional
fluctuations are then
\begin{multline} \label{Eq:GravitonFluc}
 \frac{\Delta(\bar n^{\mu} \bar n_{\mu})^2}{(n^{\mu}n_{\mu})^2} =
\frac{1}{4}\sum_{j=z_0}^{z_1} e^{2\rho_j} \Bigg|
\int_{j\varepsilon_j}^{(j+1)\varepsilon_j} d\omega
\frac{\ell_P\left(1+3\cos2\theta\right)}{324M\sqrt{\pi \varepsilon_j \tilde L
\omega}}\Big[\frac{T(\omega)e^{-i\omega(v_0+\delta_j)}}{(i+4M\omega)} \\ \times
\left(11800i + 34321M\omega  -39783iM^2\omega^2+14348M^3\omega^3\right) -
\mbox{H.C.}\Big] \Bigg|^2.
\end{multline}

Once again, one must consider that the expansion in $(V-V_0)$ is really an
expansion in $2M\omega(V-V_0)$. In the limit $\omega \to 0$, the transmission
coefficient is $T(\omega) \approx C_{\ell}(2i\omega M)^{\ell+1}$, 
where similarly
to the scalar case, $|C_{\ell}|^2\approx 1$.  Specializing to $\ell = 2$,
$T(\omega) \approx -iC_2(2M)^3\omega^3$ and $\tilde L = \frac{32}{5}$.  Then
\begin{multline} \label{Eq:GravLowFreq}
 \frac{\Delta(\bar n^{\mu} \bar n_{\mu})^2}{(n^{\mu}n_{\mu})^2} =
\frac{5}{2}\left|C_2\right|^2\left(\frac{1475(1+3\cos2\theta)}{81}\right)^2
\sum_{j=z_0}^{z_1} \frac{e^{2\rho_j}\ell_P^2 (2M)^4}{\pi\varepsilon_j}\\ \times
\left[\int_{j\varepsilon_j}^{(j+1)\varepsilon_j} d\omega\, \omega^{5/2}\cos
\omega(v_0+\delta_j)\right]^2.
\end{multline}

When $\omega \approx (2M)^{-1}$, the transmission coefficient is of order 1. It
follows that if the wavepacket is sharply peaked near $\omega \approx
(2M)^{-1}$, then
\begin{multline} \label{Eq:GravHighFreq}
 \frac{\Delta(\bar n^{\mu} \bar n_{\mu})^2}{(n^{\mu}n_{\mu})^2} \approx
\frac{(1+3\cos2\theta)^2}{16} \\ \times \sum_{j=z_0}^{z_1}
\frac{e^{2\rho_j}\ell_P^2}
{\pi\varepsilon_j}\frac{1}{M}\left[\int_{j\varepsilon_j}^{(j+1)\varepsilon_j}
d\omega\, 21.1\sin\omega(v_0+\delta_j) + 54.9\cos\omega(v_0+\delta_j)\right]^2.
\end{multline}
The physical content of this expression will be explored in 
Sect.~\ref{sec:NewConclusions},


\section{Passive Fluctuation Model} \label{sec:ScalarField}
This model differs from the scalar graviton and graviton
scenarios in that rather than a quantization of the dynamical degrees of freedom
of the gravitational field, the space-time geometry fluctuations arise passively
through fluctuations in the stress tensor of a quantized scalar field.  The
ingoing scalar field is constructed in the same manner as presented with the
scalar graviton model.  The quantity of interest is the variance of the squared
length of the separation vector, $\Delta(\bar n^{\mu}\bar n_{\mu})^2$, given by
Eq.\ (\ref{Eq:variance}). Unlike the two previous models, the second term of
this equation is not zero for the present model.

The operator nature of $\delta {R}^{\mu}_{\phantom{\mu}\alpha\nu\beta}$ is due
to the stress tensor of the scalar field. One can use (see e.g.
Ref.~\cite{Wald:1984rg}, Eq.\ 3.2.28)
\begin{equation}
 R_{\alpha\beta\mu\nu}=C_{\alpha\beta\mu\nu} + \frac{2}{n-2}
(g_{\alpha[\mu}R_{\nu]\beta} - g_{\beta[\mu}R_{\nu]\alpha}) -
\frac{2}{(n-1)(n-2)} g_{\alpha[\mu} g_{\nu]\beta}R
\end{equation}
with
\begin{equation}
 R_{\mu\nu}=8\pi\left(T_{\mu\nu} - \frac{1}{2}g_{\mu\nu}T\right), \quad
\mbox{and} \quad R=-8\pi T
\end{equation}
to find
\begin{equation}
 \delta R_{\alpha\beta\mu\nu} =
8\pi\left[g_{\alpha[\mu}T_{\nu]\beta}-g_{\beta[\mu}T_{\nu]\alpha}- \frac{2}{3}
g_{\alpha[\mu}g_{\nu]\beta}T\right]\,.
\end{equation}
Here the perturbation of the Weyl tensor vanishes, 
$\delta C_{\alpha\beta\mu\nu} = 0$, and the number of space-time
dimensions is $n=4$.  The stress tensor for a scalar field and its trace are
\begin{equation}
 T_{\mu\nu}=\Phi_{;(\mu}\Phi_{;\nu)}-\frac{1}{2}g_{\mu\nu}g^{\sigma\rho}\Phi_{
;(\sigma}\Phi_{;\rho)}, \quad \mbox{and} \quad
T=-g^{\sigma\rho}\Phi_{;(\sigma}\Phi_{;\rho)}.
\end{equation}
It follows that
\begin{equation} \label{ScalarDeltaR}
 \delta R_{\alpha\beta\mu\nu} =
8\pi\left[g_{\alpha[\mu}\Phi_{;(\nu]}\Phi_{;\beta)} -
g_{\beta[\mu}\Phi_{;(\nu]}\Phi_{;\alpha)}
-\frac{1}{3}g_{\alpha[\mu}g_{\nu]\beta}\,g^{\sigma\rho}\Phi_{;(\sigma}\Phi_{
;\rho)} \right]
\end{equation}
where it is to be understood that the antisymmetrization proceeds first, i.e.\
\begin{equation}
 g_{\alpha[\mu}\Phi_{;(\nu]}\Phi_{;\beta)} =
\frac{1}{4}\left[g_{\alpha\mu}\left( \Phi_{;\nu}\Phi_{;\beta}
+\Phi_{;\beta}\Phi_{;\nu}\right) - g_{\alpha\nu}\left( \Phi_{;\mu}\Phi_{;\beta}
+\Phi_{;\beta}\Phi_{;\mu}\right)\right].
\end{equation}

To leading order in $(V-V_0)$,
\begin{equation}
 n^{\mu}_0 \delta_pn_{\mu} = g_{UV} \, n_0^2\int_{V_0}^V dW \int_{V_0}^W dV\,
\delta R^V_{\phantom{V}VVU}.
\end{equation}
The variance is therefore proportional to an integral of a component of the
Riemann tensor correlation function
\begin{equation}
 \Delta(\bar n^{\mu}\bar n_{\mu})^2 = 4(g_{UV}n_0^2)^2 \int_{V_0}^V dW
\int_{V_0}^W dV \int_{V_0}^V dW' \int_{V_0}^{W'} dV' \, \langle
C^{V\phantom{VVU}V}_{\phantom{V}VVU\phantom{V}VVU}(x,x')\rangle
\end{equation}
where
\begin{equation}
 \langle  C^{V\phantom{VVU}V}_{\phantom{V}VVU\phantom{V}VVU}(x,x')\rangle =
\langle \delta R^V_{\phantom{V}VVU}(x)\delta R^V_{\phantom{V}VVU}(x')\rangle -
\langle \delta R^V_{\phantom{V}VVU}(x)\rangle \langle \delta
R^V_{\phantom{V}VVU}(x')\rangle.
\end{equation}

{}From Eq.~(\ref{ScalarDeltaR}), the Riemann tensor component of interest is
\begin{equation}
 \delta R^V_{\phantom{V}VVU} = \frac{8\pi}{3}\Big[\left(\Phi_{;V}\Phi_{;U} +
\Phi_{;U}\Phi_{;V}\right)  - \frac{1}{4}
g_{UV}\left[g^{\theta\theta}\left(\Phi_{;\theta}\Phi_{;\theta}+\Phi_{;\theta}
\Phi_{;\theta}\right) +
g^{\varphi\varphi}\left(\Phi_{;\varphi}\Phi_{;\varphi}+\Phi_{;\varphi}\Phi_{
;\varphi} \right) \right] \Big].
\end{equation}
For simplicity we restrict to the case $\ell=0$, so the angular derivatives
are zero.  Expanding $\Phi$ in terms of its mode functions gives
\begin{multline}
 \delta R^V_{\phantom{V}VVU} = \frac{8\pi}{3}\sum_{j,k} \left[\left(
\psi_{j,U}\psi_{k,V} + \psi_{j,V}\psi_{k,U}\right) \hat{a}_j\hat{a}_k + \left(
\psi_{j,U}\psi^*_{k,V} + \psi_{j,V}\psi^*_{k,U}\right)
\hat{a}_j\hat{a}_k^{\dagger} \right. \\ \left. + \left( \psi^*_{j,U}\psi_{k,V} +
\psi^*_{j,V}\psi_{k,U}\right) \hat{a}_j^{\dagger}\hat{a}_k + \left(
\psi^*_{j,U}\psi^*_{k,V} + \psi^*_{j,V}\psi^*_{k,U}\right)
\hat{a}_j^{\dagger}\hat{a}_k^{\dagger} \right]. \label{eq:delR}
\end{multline}
Simplify the notation by writing
\begin{equation}
 \delta R^V_{\phantom{V}VVU} = \frac{8\pi}{3}\sum_{j,k} \left[
A(x)\hat{a}_j\hat{a}_k + B(x) \hat{a}_j\hat{a}_k^{\dagger} + B^*(x)
\hat{a}_j^{\dagger}\hat{a}_k + A^*(x) \hat{a}_j^{\dagger}
\hat{a}_k^{\dagger}\right]
\end{equation}
with  $A(x)$, $A^*(x)$, $B(x)$, and $B^*(x)$ defined by comparison
with Eq.~(\ref{eq:delR}).

To find the expectation value with respect to the multimode squeezed state
$|0,\zeta\rangle = \prod_{i=z_0}^{z_1} S(\zeta_i)|0\rangle$, use the results of
Appendix \ref{app:SqueezedStates}. We again take renormalization to correspond
to restricting the sum over modes to those occupying a squeezed state.  In the
limit of large squeeze parameter, $\rho$, one finds
\begin{equation} \label{Eq:ScalarExpectation1}
 \langle\zeta,0\vert :\delta R^V_{\phantom{V}VVU}(x): \vert 0,\zeta\rangle =
-\frac{8\pi}{3}\sum_{j,k=z_0}^{z_1} \frac{e^{\rho_j}e^{\rho_k}}{2^2}\delta_{jk}
\left[ A(x) + A^*(x) - B(x) - B^*(x) \right].
\end{equation}
and
\begin{multline}
 \langle :\delta R^V_{\phantom{V}VVU}(x)\delta R^V_{\phantom{V}VVU}(x'):\rangle
=  \left(\frac{8\pi}{3}\right)^2 \sum_{j,k=z_0}^{z_1}\sum_{r,s=z_0}^{z_1}
\frac{e^{\rho_j}e^{\rho_k}e^{\rho_r}e^{\rho_s}}{2^4} (\delta_{jr}\delta_{ks} +
\delta_{js}\delta_{kr} + \delta_{jk}\delta_{rs}) \\ \times
\left[\left(A(x)+A^*(x)-B(x)-B^*(x)\right)
\left(A(x')+A^*(x')-B(x')-B^*(x')\right)\right].
\end{multline}

In general, quartic operator products can be expanded into a sum of a
fully normal ordered part, a cross term, and a vacuum part. (See, for
example, Ref.~\cite{Ford:2000vm}.) Our procedure of restricting the sum to those
modes which lie in the range of squeezing is, in the limit of large
squeeze parameter,  equivalent to retaining only the fully normal
ordered part.  To see this, consider the difference between
one of the terms above and its normal ordered version.  As a concrete example,
consider 
\begin{equation}
\langle\zeta,0\vert \hat{a}_j\hat{a}^{\dagger}_k\hat{a}_r\hat{a}_s \vert
0,\zeta\rangle - (\langle\zeta,0\vert
\hat{a}^{\dagger}_k\hat{a}_j\hat{a}_r\hat{a}_s \vert 0,\zeta\rangle +
\delta_{jk}\langle\zeta,0\vert\hat{a}_r\hat{a}_s\vert 0,\zeta\rangle).
\end{equation}
{}From equations (\ref{Expaadagaa}), (\ref{Expadagaaa}), and (\ref{Expaa}) it is
clear that this is proportional to the subdominant term $e^{2\rho}$ (compared to
$e^{4\rho}$).  Thus,  in the limit of large squeeze parameter,
restricting the sum over modes to those which lie in the range of squeezing
corresponds to taking the fully normal ordered part. Furthermore, the
cross and vacuum terms which have been neglected are sub-dominant in
this limit.

 The
$\delta_{jk}\delta_{rs}$ term is the same as  $\langle:\delta
R^V_{\phantom{V}VVU}(x):\rangle\langle:\delta R^V_{\phantom{V}VVU}(x'):\rangle$,
which cancels to leave
\begin{multline}
 \langle C^{V\phantom{VVU}V}_{\phantom{V}VVU\phantom{V}VVU}(x,x')\rangle =
\frac{1}{2^4}\left(\frac{8\pi}{3}\right)^2
\sum_{j,k=z_0}^{z_1}\sum_{r,s=z_0}^{z_1}
e^{\rho_j}e^{\rho_k}e^{\rho_r}e^{\rho_s} (\delta_{jr}\delta_{ks} +
\delta_{js}\delta_{kr}) \\ \times \left[\left(A(x)+A^*(x)-B(x)-B^*(x)\right)
\left(A(x')+A^*(x')-B(x')-B^*(x')\right)\right].
\end{multline}
Using the definitions of $A(x)$ and $B(x)$, noting that the sums extend over the
same range, and using the fact that the Kronecker deltas act symmetrically, one
may show that
\begin{multline}
 \langle C^{V\phantom{VVU}V}_{\phantom{V}VVU\phantom{V}VVU}(x,x')\rangle =
4\left(\frac{8\pi}{3}\right)^2 \sum_{j,k=z_0}^{z_1} e^{2\rho_j}e^{2\rho_k} 
\left[\left(\mathrm{Im}\psi_j(x)\right)_{,V}\left(\mathrm{Im}\psi_k(x)\right)_{,
U}\right] \\ \times
\left[\left(\mathrm{Im}\psi_j(x')\right)_{,V'}\left(\mathrm{Im}
\psi_k(x')\right)_{,U'} \right].
\end{multline} 
Since $\langle C^{V\phantom{VVU}V}_{\phantom{V}VVU\phantom{V}VVU}(x,x')\rangle$
is a product of a function of $x$ with a function of $x'$, the
integral over $x$ and $x'$ of the product is the product of the integrals such
that
\begin{multline}
 \int_{V_0}^{V} dW \int_{V_0}^{W} dV \int_{V_0}^{V} dW' \int_{V_0}^{W'} dV' \,
\langle C^{V\phantom{VVU}V}_{\phantom{V}VVU\phantom{V}VVU}(x,x')\rangle = \\
4\left(\frac{8\pi}{3}\right)^2 \sum_{j,k=z_0}^{z_1} e^{2\rho_j}e^{2\rho_k}
\left| \int_{V_0}^{V} dW \int_{V_0}^{W} dV
\left(\mathrm{Im}\psi_j(x)\right)_{,V}\left(\mathrm{Im}\psi_k(x)\right)_{,U}
\right|^2.
\end{multline}
Near the horizon, the derivatives of $\psi$ with respect to $U$ and $V$ are
straightforwardly calculated.  One subsequently finds
\begin{multline}
 \left(\mathrm{Im}(\psi_j)\right)_{,V}\left(\mathrm{Im}(\psi_k)\right)_{,U} =
\frac{ie^{-1}}{8\pi^2M\sqrt{\varepsilon_j\varepsilon_k}}
\int_{j\varepsilon_j}^{(j+1)\varepsilon_j}d\omega
\int_{k\varepsilon_k}^{(k+1)\varepsilon_k}d\omega'\,
\left(\frac{\omega}{\omega'}\right)^{\tfrac{1}{2}} \\ \times
\Big[F(\omega)F(\omega') e^{-i(\omega\delta_j+\omega'\delta_k)}
e^{-iv(\omega+\omega')}  \\ - F(\omega)F^*(\omega')
e^{-i(\omega\delta_j-\omega'\delta_k)} e^{-iv(\omega-\omega')} + H.C. \Big].
\end{multline}
The integration over $V$ of
$\left(\mathrm{Im}(\psi_j)\right)_{,V}\left(\mathrm{Im}(\psi_k)\right)_{,U}$
reduces to calculating
\begin{equation}
 \int_{V_0}^V dW \int_{V_0}^W dV\, e^{\mp iv(\omega\mp\omega')} = \int_{V_0}^V
dW \int_{V_0}^W dV\, V^{\mp 4iM(\omega\mp\omega')} \approx \frac{1}{2} e^{\mp i
v_0(\omega\mp\omega')}(V-V_0)^2 \, ,
\end{equation}
where the solution has been expanded in powers of $(V-V_0)$.  Inserting the
appropriate powers of the Planck length, it follows that the fractional
fluctuations are characterized by
\begin{multline} \label{Eq:ScalarFluc}
 \frac{\Delta(\bar n^{\mu}\bar n_{\mu})^2}{(n^{\mu}n_{\mu})^2} =
\sum_{j,k=z_0}^{z_1} \left|\frac{ e^{\rho_j}e^{\rho_k} \ell_P^2}{3\pi M}
\int_{j\varepsilon_j}^{(j+1)\varepsilon_j}d\omega\,
\sqrt{\frac{\omega}{\varepsilon_j}}\left(F(\omega)e^{-i\omega(v_0+\delta_j)} -
F^*(\omega)e^{i\omega(v_0+\delta_j)}\right) \right. \\ \left. \times
\int_{k\varepsilon_k}^{(k+1)\varepsilon_k}d\omega'\,
\frac{1}{\sqrt{\omega'\varepsilon_k}} \left(F(\omega')e^{-i\omega(v_0+\delta_k)}
- F^*(\omega')e^{i\omega(v_0+\delta_k)}\right)\right|^2 \,.
\end{multline}

Once again, one must consider that the expansion in $(V-V_0)$ is really an
expansion in $2M\omega(V-V_0)$. In the limit $\omega \to 0$, the transmission
coefficient is  again $F(\omega) \sim B_{\ell}(2i\omega M)^{\ell+1}$. 
Specializing to $\ell = 0$, $F(\omega) \sim 2iB_0M\omega$, and the fractional
fluctuations are then
\begin{multline} \label{Eq:SFLowFreq}
 \frac{\Delta(\bar n^{\mu}\bar n_{\mu})^2}{(n^{\mu}n_{\mu})^2} =
\left(\frac{16}{3}\right)^2\left| B_0\right|^4 \sum_{j,k=z_0}^{z_1}
\frac{e^{2\rho_j} \ell_P^2 M}{\pi\varepsilon_j} \left[
\int_{j\varepsilon_j}^{(j+1)\varepsilon_j}d\omega\,
\omega^{3/2}\cos\omega(v_0+\delta_j)\right]^2  \\ \times \frac{e^{2\rho_k}
\ell_P^2 M}{\pi\varepsilon_k} \left[
\int_{k\varepsilon_k}^{(k+1)\varepsilon_k}d\omega'\,
(\omega')^{1/2}\cos\omega'(v_0+\delta_k)\right]^2.
\end{multline}
Near $\omega \approx (2M)^{-1}$, the transmission coefficient is of order 1. It
follows that if the wavepacket is sharply peaked near $\omega \approx
(2M)^{-1}$, then
\begin{multline} \label{Eq:SFHighFreq}
 \frac{\Delta(\bar n^{\mu}\bar n_{\mu})^2}{(n^{\mu}n_{\mu})^2} =
\left(\frac{4}{3}\right)^2 \sum_{j,k=z_0}^{z_1} \frac{e^{2\rho_j} \ell_P^2}{\pi
\varepsilon_j M} \left[
\int_{j\varepsilon_j}^{(j+1)\varepsilon_j}d\omega\,\sin\omega(v_0+\delta_j)
\right]^2 \\ \times \frac{e^{2\rho_k} \ell_P^2}{\pi \varepsilon_k M} \left[
\int_{k\varepsilon_k}^{(k+1)\varepsilon_k}d\omega'\,
\sin\omega'(v_0+\delta_k)\right]^2.
\end{multline}


\section{Discussion} \label{sec:NewDiscussion}
\subsection{Summary of Results}

Consider the results for the three different models -- scalar graviton,
graviton, and  stress tensor induced fluctuations.  There are two
limits of interest, $\omega\to 0$ and $\omega \approx (2M)^{-1}$.

\subsubsection*{Case 1: $\omega\ll 1/M$}
In the low frequency limit, $\omega\to 0$, we found Eqs.\ (\ref{Eq:SGLowFreq}),
(\ref{Eq:GravLowFreq}), and (\ref{Eq:SFLowFreq}) for the scalar graviton,
graviton, and passive fluctuation models, respectively.
The solutions to these integrals may be expressed in terms of incomplete gamma
functions, but it is not necessary to invoke the machinery of incomplete gamma
functions to get an idea of the general behavior of the fluctuations.  To
simplify the discussion, let us set $\delta_j=0$.  This may be assumed without
loss of generality and is equivalent to assuming $n=0$ in 
Eq.~(\ref{Eq:WavePacket}).  In this limit we may use the small 
angle approximation
to set $\cos(\omega v_0) \approx 1$.  Furthermore, for a sharply peaked
wavepacket, we may assume the integrand is approximately constant so that
\begin{equation}
 \int d\omega \, \omega^x \cos(\omega v_0) \approx \omega^x\Delta\omega.
\end{equation}
Ignoring the numerical factors and recognizing that $\varepsilon_j =
\Delta\omega$, the general behavior of the fractional fluctuations is

\begin{equation} \label{Eq:LowFreqFlucs}
 \frac{\Delta(\bar n^{\mu}\bar n_{\mu})^2}{(n^{\mu}n_{\mu})^2} \approx
 \begin{cases}
  \sum_j \frac{1}{\pi}e^{2\rho_j}\ell_P^2(2M\omega)^{2\ell} \omega \Delta\omega,
& \mbox{Scalar Graviton}, \\
  \sum_j \frac{1}{\pi} e^{2\rho_j}\ell_P^2(2M\omega)^4 \omega \Delta\omega, &
\mbox{Graviton}, \\
  \left( \sum_j \frac{1}{\pi} e^{2\rho_j}\ell_P^2 (2M\omega) \omega
\Delta\omega\right)^2, & \mbox{Passive}.
 \end{cases}
\end{equation}

\subsubsection*{Case 2: $\omega \approx (2M)^{-1}$}
For $\omega \approx (2M)^{-1}$, on the other hand, the fractional 
fluctuations were
given by Eqs.\ (\ref{Eq:SGHighFreq}), (\ref{Eq:GravHighFreq}), and 
(\ref{Eq:SFHighFreq}).
Since the wavepacket is assumed to be sharply peaked in $\omega$, then
$\Delta\omega = \varepsilon_j \ll \omega$ while $j$ is large.  
Thus we may use the
small angle approximation, $\cos(\varepsilon_j v_0) \approx 1$ and
$\sin(\varepsilon_j v_0) \approx \varepsilon_j v_0$.  Furthermore, as an order
of magnitude estimate, we may assume $\sin(j\varepsilon_j v_0)$ is of order 1 so
that
\begin{equation}
 \int_{j\varepsilon_j}^{(j+1)\varepsilon_j} d\omega \sin(\omega v_0) \approx 
\varepsilon_j = \Delta\omega,
\end{equation}
and similarly
\begin{equation}
 \int_{j\varepsilon_j}^{(j+1)\varepsilon_j} d\omega \cos(\omega v_0) \approx 
\varepsilon_j = \Delta\omega.
\end{equation}

The general behavior of the fractional fluctuation in the limit $\omega \approx
(2M)^{-1}$ is now
\begin{equation} \label{Eq:HighFreqFlucs}
 \frac{\Delta(\bar n^{\mu}\bar n_{\mu})^2}{(n^{\mu}n_{\mu})^2} \approx
 \begin{cases}
  \sum_j \frac{e^{2\rho_j}\ell_P^2 \Delta\omega}{\pi M}, & \mbox{Scalar
Graviton and Graviton}, \\
  \left( \sum_j \frac{e^{2\rho_j}\ell_P^2 \Delta\omega}{\pi M} \right)^2, &
\mbox{Passive}.
 \end{cases}
\end{equation}
Consider for a moment the limit of low squeezing, $\rho\to 1$.  Although our
results have been derived for $\rho \gg 1$, the order of magnitude behavior
should still correspond to what we have derived since $\sinh\rho$ and
$\cosh\rho$ are of order one for small $\rho$.  In this case the active
fluctuations behave as $\ell_P^2\Delta\omega/M$, which is in agreement with the
results of Ford and Svaiter \cite{Ford:1997zb} where they find fluctuations in
the proper time of an infalling observer become of order one for $M\sim \ell_p$.

The scalar graviton and graviton models describe the active
fluctuations, while the scalar field stress tensor fluctuations are 
 passive fluctuations.  Specializing $\ell = 2$ in the scalar graviton
 model, the general
behavior is identical to that of the graviton model.  In light of Eqs.~
(\ref{Eq:ScalarGravFluc}), (\ref{Eq:GravitonFluc}), and (\ref{Eq:ScalarFluc})
one sees that Eqs.~(\ref{Eq:LowFreqFlucs}), and (\ref{Eq:HighFreqFlucs})
become
\begin{equation} \label{Eq:GeneralFlucs}
 \frac{\Delta(\bar n^{\mu}\bar n_{\mu})^2}{(n^{\mu}n_{\mu})^2} \sim
 \begin{cases}
  \sum_j \frac{e^{2\rho}\ell_P^2 \vert F(\omega)\vert^2
\Delta\omega}{M(M\omega)}, & \mbox{Active}, \\
  \left(\sum_j \frac{e^{2\rho}\ell_P^2 \vert F(\omega)\vert^2
\Delta\omega}{M}\right)^2, & \mbox{Passive}.
 \end{cases}
\end{equation}

Although we have specialized to $\ell = 0$ for the passive fluctuations, it is
not difficult to generalize to the more general case of arbitrary $\ell>0$.  For
$\ell>0$ there are some additional derivatives of $\Phi$ with respect to the
angular variables, but this will not alter the order of magnitude behavior in
Eq.\ (\ref{Eq:GeneralFlucs}).

There are some generic features common to both active and passive fluctuations. 
As $\omega\to 0$, the fractional fluctuations are suppressed by some power of
$\omega$.  They are further suppressed by a factor of $\Delta\omega$, which has
been assumed small, and some powers of the Planck length.  Notice that the
passive fluctuations are more heavily suppressed since the passive fluctuations
are proportional to $\ell_P^4$ whereas the active fluctuations are proportional
to $\ell_P^2$.  The suppression in the low frequency limit is expected, since
the effective potential barrier efficiently prevents low frequency modes from
reaching the horizon.  However, it seems these suppressions may be overcome by
arbitrarily increasing the squeeze parameter.

For $\omega \approx (2M)^{-1}$ the similarities between the active and passive
fluctuations are even more striking.  In this case, the fluctuations are again
suppressed by the same powers of the Planck length and the width of the
wavepacket, but are additionally suppressed by the black hole mass.  One would
expect this term to become important once the black hole evaporates to the
Planck mass. Again, by arbitrarily increasing the squeeze parameter it is
possible to overcome the various suppressions.

\subsection{Semiclassical Restriction on Squeezing}
Since it appears that the fluctuations may become arbitrarily large by
unboundedly increasing $\rho$, we should investigate whether there is an upper
bound on $\rho$.  After all, a squeezed vacuum state is not necessarily devoid
of particle content.  From the discussion in Appendix \ref{app:SqueezedStates}
it follows that for a single mode squeezed vacuum state, $|0,\zeta\rangle =
S(\zeta)|0\rangle$, the expectation value for the number of particles is
\begin{equation}
 \langle \zeta,0|\hat N|0,\zeta\rangle = \langle 0| S^{\dagger}(\zeta) \hat
a^{\dagger} \hat a S(\zeta)|0\rangle = \sinh^2 \rho \to e^{2\rho}, \, \rho\gg 1.
\end{equation}
This means that increasing the squeeze parameter increases the mean energy
density.  If the energy density grows too large, then the semiclassical
approximation in which backreaction is ignored fails.

Consider the passive fluctuation model.  
>From the semiclassical Einstein equation,
these calculations should remain valid as long as
\begin{equation}
 \langle \delta R^{\alpha}_{\phantom{\alpha}\beta\mu\nu}\rangle \ll
R^{\alpha}_{\phantom{\alpha}\beta\mu\nu}.
\end{equation}
Working in null Kruskal coordinates, a typical component of the background
Riemann tensor is
\begin{equation}
 R^V_{\phantom{V}VVU} = \frac{16M^3}{UVr^3}\left(1-\frac{2M}{r}\right),
\end{equation}
which, since $\left(1-\frac{2M}{r}\right)\approx -UV$ near the horizon, is
approximately $-2e^{-1}$.  However, $\delta R^V_{\phantom{V}VVU}$ is precisely
the quantity that was calculated for the passive fluctuation model.  Using Eq.\
(\ref{Eq:ScalarExpectation1}) it follows that
\begin{equation}
 \sum_j \frac{e^{2\rho_j}\ell_P^2 \vert F(\omega)\vert^2 \Delta\omega}{M} \ll
 1.
\end{equation}

Consider next the scalar graviton model.  In this case $\delta
R_{\alpha\beta\mu\nu}$ is linear in the field, and therefore $\langle \delta
R_{\alpha\beta\mu\nu}\rangle = 0$.  One could look for second order
perturbations $\delta R^{(2)}_{\alpha\beta\mu\nu}\propto (h_{\mu\nu})^2$ that
would be quadratic in the field.  Instead, we consider
\begin{equation}
 \langle \delta R_{\alpha\beta\mu\nu} \delta R_{\alpha\beta\mu\nu} \rangle \ll
R_{\alpha\beta\mu\nu} R_{\alpha\beta\mu\nu}
\end{equation}
to be a good indicator of whether the semiclassical treatment is valid for the
perturbation.  The quantity of interest is 
\begin{equation}
 \langle \delta R^V_{\phantom{V}VVU} \delta R^V_{\phantom{V}VVU} \rangle =
\langle \Phi_{,VU}\Phi_{,VU}\rangle
\end{equation}
and the calculations proceed as in Sec. \ref{sec:ScalarGraviton}.  The result
for $\langle \Phi_{,VU}\Phi_{,VU}\rangle$ is precisely the same as that found
for $\Delta(\bar n^{\mu}\bar n_{\mu})^2/(n^{\mu}n_{\mu})^2$ in Eq.\
(\ref{Eq:ScalarGravFluc}) and we therefore have
\begin{equation} \label{Eq:SGRestriction}
 \sum_j \frac{e^{2\rho}\ell_P^2 \vert F(\omega)\vert^2 \Delta\omega}{M(M\omega)}
\ll 1\, .
\end{equation}
For the graviton model, one must again turn to a computer algebra system to
calculate $\langle \delta R^V_{\phantom{V}VVU} \delta R^V_{\phantom{V}VVU}
\rangle$.  The result agrees with Eq.\ (\ref{Eq:SGRestriction}).
In the process of seeking to place an upper bound on the results of Eq.
(\ref{Eq:GeneralFlucs}) via a restriction on the squeeze parameter, $\rho$, a
restriction on the results of Eq. (\ref{Eq:GeneralFlucs}) themselves has been
found.  It is therefore sufficient to proceed with this restriction on the
general result.

\subsection{Analysis of Results}
For the sake of an order of magnitude estimate of the fractional fluctuations,
assume the extremal cases where
\begin{equation}
 \sum_j \frac{e^{2\rho_j}\ell_P^2 \vert F(\omega)\vert^2 \Delta\omega}{M} 
\approx 1
\end{equation}
for the active fluctuations, and
\begin{equation}
 \sum_j \frac{e^{2\rho_j}\ell_P^2 \vert
F(\omega)\vert^2 \Delta\omega}{M(M\omega)} \sim 1
\end{equation}
for the passive fluctuations.
Using this restriction, the fractional fluctuations become
\begin{equation} \label{Eq:FlucResults}
 \frac{\Delta(\bar n^{\mu}\bar n_{\mu})^2}{(n^{\mu}n_{\mu})^2} \lesssim
 \begin{cases}
  1, & \mbox{Active}, \\
  1, & \mbox{Passive}.
 \end{cases}
\end{equation}

At first glance, it appeared that the fractional fluctuations described in Eq.\
(\ref{Eq:GeneralFlucs}) could be made arbitrarily large by increasing the
squeeze parameter, a quite surprising result. Since $\bar n^{\mu} \bar
n_{\mu}$ reflects the change in geodesic deviation from the Schwarzschild
background, restricting to perturbations that are in some sense small would lead
one to expect that $\bar n^{\mu} \bar n_{\mu}$ should also be small and not
deviate very much from the background value, $n^{\mu} n_{\mu}$ (equivalently the
expectation value of the fluctuating quantity).  Indeed, after imposing
restrictions on the amount of allowed squeezing by requiring the induced
curvature to be small compared to the background, we find the fluctuations to be
no more than of order one.
Nonetheless, fractional fluctuations of order unity in $\Delta(\bar
n^{\mu}\bar n_{\mu})^2/(n^{\mu}n_{\mu})^2$ have the potential to
dramatically alter the outgoing radiation.

\subsection{Implications for Hawking Radiation}
Do the space-time fluctuations implied by Eq.\ (\ref{Eq:FlucResults}) have any
significant effect on Hawking radiation?  Fluctuations of the vector separating
the horizon from a nearby outgoing null geodesic have been studied here
precisely because of this vector's importance to the Hawking effect derivation. 
However, it is the $U$ component of the separation vector that is crucial to
Hawking's derivation.  In fact, $\delta_p n^U \equiv 0$ by the 
symmetry properties 
of the Riemann tensor. Thus all of the contributions to the quantity
${\Delta(\bar n^{\mu}\bar n_{\mu})^2}/{(n^{\mu}n_{\mu})^2}$ come from
the $V$-component of $\bar n^{\mu}$.

Fluctuations in $n^V$ indicate that an outgoing null geodesic
may take either a longer or shorter than average affine time in reaching some
distance from the black hole.  In this sense the results tend to agree with the
heuristic picture of Ford and Svaiter \cite{Ford:1997zb}.
This would correspond to an uncertainty in the knowledge of which wavepacket was
under consideration.  Recall that in the derivation of the Hawking effect the
ingoing wavepackets were controlled by an integer, $n$, which allowed for
successive ingoing wavepackets.  A fluctuation in $n^V$ would mean that the
ordering of these successive wavepackets could be disrupted.  This would not
have any observable effect on the outgoing radiation if the ingoing state is the
vacuum.
But it is also possible to have stimulated emission in addition to the thermal
flux.  This would be caused by particles that are initially present during
collapse and has been studied by Wald \cite{Wald:1976ka}.  In this case the
stimulated emission only occurs at early times while the late time behavior
remains thermal.  Fluctuations in $n^{V}$ could allow for fluctuations in the
arrival time of particles originating from this stimulated flux.

Note that effects near the horizon do not necessarily translate to
effects observed at $\scri^+$.  Indeed, we were restricted to considering
fluctuations only along a short segment of the outgoing geodesic.  In order to
discuss observations made by a distant observer, one would really have to be
able to follow the evolution of the separation vector all the way out to the
observer.  In doing so one would see whether the fluctuations integrate to
produce a large effect or not.  The presence of the effective potential barrier
prevents us from tracking the long term evolution of the fluctuations and so at
this point discussing observations of a distant observer is merely speculation. 
Future research may help resolve this issue.  In particular, particle creation
near a moving mirror has been shown to be analogous to the Hawking effect for
black holes, with an identical thermal spectrum obtained for specific mirror
trajectories \cite{Davies:1976hi,Davies:1977yv,Ford:1982ct}.  Further insight
may be obtained into the horizon fluctuations of black holes by considering
fluctuations in the trajectory of the mirror.  The benefit of considering a
moving mirror is that it lacks the complicated potential barrier of a black
hole, allowing one to integrate out to an observer.

\section{Summary and Conclusions} \label{sec:NewConclusions}
Deviation of outgoing null geodesics in a fluctuating Schwarzschild geometry
have been considered.  The reference geodesic was taken to be that outgoing null
geodesic which generates the future horizon in the average (Schwarzschild)
space-time.  Fluctuations of the space-time were induced both actively and
passively.  The active fluctuations were first described by a 
scalar graviton model,
consisting of a conformal transformation of the Schwarzschild metric where the
conformal factor is a quantized scalar field, and then by 
a graviton model that is
constructed as a linear quantum tensor field where the mode functions are taken
to be the classically allowed even parity Schwarzschild perturbations.  
The passive
fluctuations arise from stress tensor fluctuations of a free scalar field.
 In all models the ingoing perturbation is taken to occupy a multimode 
squeezed vacuum state. The modes in question are wavepackets being
sent into the black hole after the collapse process.

For all models, fractional fluctuations of the quantity $\bar n^{\mu}\bar
n_{\mu}$ were calculated. The vacuum level fluctuations are very small
for large black holes, being of order $(\ell_p/M)^2$ for active
fluctuations and  of order $(\ell_p/M)^4$ for passive ones. However,
 the fluctuations could be boosted by increasing the
amount of squeezing.  An upper bound on the squeeze parameter was imposed by
restricting the energy density of the ingoing field to fall within the allowed
limits of semiclassical theory ignoring backreaction.  This upper bound was then
used to estimate an order of magnitude for the fractional fluctuations,
$\Delta(\bar n^{\mu}\bar n_{\mu})^2/(n^{\mu}n_{\mu})^2$.  It was found
that the fractional fluctuations could in principle become of order unity.

Such large fluctuations would seem at first sight to have a dramatic
effect upon the outgoing modes which carry the created
particles. However, the fluctuations come from the $V$-component of
$n^\mu$, not the $U$-component. This implies that the primary effect of
these fluctuations is on the time delay of individual wavepackets,
which would only be observable in stimulated emission, but not in
spontaneous emission arising when the quantum state is the in-vacuum. 
This suggests that the thermal nature of the Hawking radiation is quite
robust and is not altered by the type of enhanced fluctuations we have
studied.

\begin{acknowledgments}
This work was supported in part by the National
Science Foundation under Grant PHY-0555754.
\end{acknowledgments}

\appendix
\section{Scalar Field Normalization} \label{app:ScalarNormalization}
The scalar field normalization is found via the Klein-Gordon norm
\begin{equation}
 \langle\psi_{jn},\psi_{j'n'}\rangle = i\int_S d\Sigma_{\mu}\left(
\psi_{jn}^*\stackrel{\leftrightarrow\phantom{\mu}}{\nabla^{\mu}}\psi_{j'n'}
\right)= \delta_{\ell\ell'}\delta_{mm'}.
\end{equation}
The surface over which the integral is performed is naturally $\scri^-$, past
null infinity, and the coordinates appropriate in this region are ingoing
Eddington-Finkelstein coordinates $(v,r,\theta,\varphi)$.  In these coordinates,
$\scri^-$ is a three-surface isomorphic to $S^2\times \mathbb{R}$, or a sphere
at infinity further parameterized by $v$.  The normal to this surface is
therefore in the direction of ${\partial}/{\partial r}$ and the preceding
integral is
\[
 i\int_{\scri^-}d\Sigma_{\mu}\,\left( \psi_{jn}^*
\stackrel{\leftrightarrow\phantom{\mu}}{\nabla^{\mu}} \psi_{j'n'}\right) =
i\int_{S^2} r^2d\Omega\int_{-\infty}^{\infty} dv\, \left(
\psi_{jn}^*\stackrel{\leftrightarrow\phantom{r}}{\nabla^{r}}\psi_{j'n'}\right).
\]
The field $\psi_{jn}$ is a scalar field, for which
$\nabla^{\mu}=\partial^{\mu}$.  The metric for ingoing Eddington-Finkelstein
coordinates is off diagonal, and the $(v,r)$ sector is
\begin{equation}
 g_{\mu\nu} = \left(
  \begin{matrix}
   -\left(1-\frac{2M}{r}\right) & 1 \\
   1 & 0
  \end{matrix} \right)
  \qquad
 g^{\mu\nu} = \left(
  \begin{matrix}
   0 & 1 \\
   1 & \left(1-\frac{2M}{r}\right)
  \end{matrix} \right).
\end{equation}
{}From this one finds
\begin{equation}
 \partial^r = g^{rv}\partial_v+g^{rr}\partial_r =
\partial_v+\left(1-\frac{2M}{r}\right)\partial_r
\end{equation}
and may proceed to calculate
\begin{equation}
 i\int_{S^2}r^2d\Omega\int_{-\infty}^{\infty}dv \, \psi_{jn}^*
\left[\stackrel{\leftrightarrow\phantom{v}}{\partial_v}+\left(1-\tfrac{2M}{r}
\right)
\stackrel{\leftrightarrow\phantom{r}}{\partial_r}\right]\psi_{j'n'}\,.
\label{eq:norm}
\end{equation}
The calculations are straightforward, but one must bear in mind that while the
advanced time is implicitly defined in terms of $r$, when taking the derivative
with respect to $v$ or $r$, the other variable is held fixed.  Additionally, the
$r$ dependence is the same for both $\psi_{jn}^*$ and $\psi_{j'n'}$ so the $r$
derivative terms cancel.  Examining the norm in the asymptotically flat region
$r_*\rightarrow r \rightarrow \infty$, Eq.~(\ref{eq:norm}) becomes
\begin{multline}
 \frac{i}{2\pi\varepsilon_j}\int d\Omega\, Y_{L m}^*(\theta,\varphi)Y_{L'
m'}(\theta,\varphi) \\ 
\times \int_{j\varepsilon_j}^{(j+1)\varepsilon_j} d\omega
\int_{j\varepsilon_j}^{(j+1)\varepsilon_j} d\omega' e^{2\pi
i(\omega-\omega')/\varepsilon_j}\frac{F^*_{\omega}(r)F_{\omega'}(r)}{\sqrt{
\omega\omega'}} \int_{-\infty}^{\infty} dv \, (e^{i\omega v}
\stackrel{\leftrightarrow}{\partial_v} e^{-i\omega' v}).
\end{multline}

{}From the normalization condition of the spherical harmonics, the integral over
the sphere gives a product of delta functions, $\delta_{\ell\ell'}\delta_{mm'}$.
 The function $F(r)$ was included in the definition of the ingoing null field to
reflect the presence of an effective potential barrier due to the space-time
curvature, but $F(\scri^-)=1$ where $r\to\infty$ at $\scri^-$.  Lastly,
$e^{i\omega v} \stackrel{\leftrightarrow}{\partial_v} e^{-i\omega' v} =
-i(\omega+\omega') e^{i(\omega-\omega')v}$ so the integration with respect to
$v$ gives a delta function $\pi\delta(\omega-\omega')$. This reduces the
normalization condition at infinity to 
\begin{equation}
 \langle\psi_{jn},\psi_{j'n'}\rangle  =
(\varepsilon_j)^{-1}\delta_{\ell\ell'}\delta_{mm'}\int_{j\varepsilon_j}^{
(j+1)\varepsilon_j} d\omega =\delta_{\ell\ell'}\delta_{mm'}.
\end{equation}
So as given, $\psi_{jn}$ is properly normalized.


\section{Graviton Mode Normalization} \label{app:GravNormalization}
Some care is required when fixing the normalization of the ingoing graviton
wavepackets.  We choose to fix the normalization in the asymptotically
flat-space region of $r_*\to\infty$ by setting the vacuum energy of each mode to
$\tfrac{1}{2}\omega$.  One would like to first calculate the effective energy
density of the wave, integrate over some volume, and set the result to
$\tfrac{1}{2}\omega$.  The typical approach in calculating the energy density
would be to use Eq.~35.70 in Ref. \cite{Misner:1974qy}, which
relates the effective energy density of a gravitational wave to derivatives of
the perturbation.  The problem is that in the Regge-Wheeler gauge, where
calculations are most easily performed, $\Psi_{(j)\mu\nu}$ is not well-behaved
at infinity.  In fact, $H_0(r)$, $H_1(r)$, and $H_2(r)$ all grow linearly with $r$ as
$r_*\to\infty$.  While Eq.~35.70 of Ref. \cite{Misner:1974qy} is gauge
invariant, it implicitly involves some averaging and assumptions about covariant
divergences at infinity that do not hold in the Regge-Wheeler gauge. 
Alternatively, it should be possible to transform the perturbations to the
radiation gauge, where they are well-behaved, but the calculations become more
difficult in the radiation gauge.  Instead, we choose to take a somewhat
circuitous route by first calculating the (gauge invariant) Riemann tensor at
infinity.  We then transform the Riemann tensor to Cartesian coordinates and
make use of a special relation between the Riemann tensor and gravitational
waves in the Transverse Tracefree (TT) gauge.  Calculating the effective stress
tensor of a gravity wave in the TT gauge is then  straightforward using
\begin{equation}
 T_{\mu\nu}^{eff}=\tfrac{1}{32\pi}\langle \Psi_{(j)ab,\mu}^{TT}
(\Psi_{(j),\nu}^{\phantom{(j)}ab(TT)})^*+(\Psi_{(j)ab,\mu}^{TT})^*
\Psi_{(j),\nu}^{\phantom{(j)}ab(TT)}\rangle.
\end{equation}
Here the brackets indicate a spatial average over several wavelengths.  The
components of $\Psi_{(j)\mu\nu}^{TT}$ have a simple relationship to components
of the Riemann tensor,  $\Psi_{(j)kl,00}^{TT}=-2R_{0 k0 l}$.  Since
$\Psi_{(j)\mu\nu}$ has a simple sinusoidal time dependence, it follows that
$\Psi^{TT}_{(j)kl,0}=-\imath\omega \Psi^{TT}_{(j)kl}$ and so
$\Psi^{TT}_{(j)kl}=2\omega^{-2}R_{0k0l}$.  In the asymptotically flat space at
infinity,
\begin{equation}
 R_{\alpha\beta\mu\nu}=\tfrac{1}{2}(\Psi_{(j)\alpha\nu,\beta\mu}-\Psi_{
(j)\alpha\mu,\beta\nu}+
\Psi_{(j)\beta\mu,\alpha\nu}-\Psi_{(j)\beta\nu,\alpha\mu}).
\end{equation}
Let $J^{\mu}_{\nu'}$ be the transformation matrix from spherical to Cartesian
coordinates.  Then in Cartesian coordinates
\begin{equation}
 R_{0\beta' 0\nu'}=\delta^{\alpha'}_0\delta^{\mu'}_0
R_{\alpha'\beta'\mu'\nu'}=\delta^{\alpha'}_0\delta^{\mu'}_0
J^{\gamma}_{\alpha'}J^{\lambda}_{\beta'}J^{\rho}_{\mu'}J^{\sigma}_{\nu'}R_{
\gamma\lambda\rho\sigma}=J^{\lambda}_{\beta'}J^{\sigma}_{\nu'}R_{0\lambda 0
\sigma}
\end{equation}
while the TT metric perturbation is
$\Psi^{TT}_{(j)kl}=-\omega^{-2}J^{\lambda}_{k}J^{\sigma}_{l}R_{0\lambda 0
\sigma}$.  Carrying out the above calculations, one finds 
\begin{multline}
T^{eff}_{00}=\varepsilon^{-1}\left(P^2_{\ell}
(\cos\theta)\left(1+(1-\ell(\ell+1))^2\right) \right. \\ - \left. 2\cos\theta
P'_{\ell}(\cos\theta)\left(\sin^2\theta
P^{\prime\prime}_{\ell}(\cos\theta)-\cos\theta
P'_{\ell}(\cos\theta)\right)\right) \\ \times \int d\omega \int d\omega' \frac{C
C'\omega\omega'}{16\pi r^2} \cos\left((r+t+\delta)(\omega-\omega')\right).
\end{multline}
Next, for ingoing null waves, the total energy on the surface of a sphere is the
same as the flux through the sphere.  The flux through the surface of a sphere
is therefore
\begin{multline}
 \Phi = \int_{\Sigma}T^{eff}_{00} r^2 d\Omega = \varepsilon^{-1} 2\pi
\int_{-\pi}^{\pi} \sin\theta d\theta
\left(P^2_{\ell}(\cos\theta)\left(1+(1-\ell(\ell+1))^2\right) \right. - \\
\left. 2\cos\theta P'_{\ell}(\cos\theta)\left(\sin^2\theta
P^{\prime\prime}_{\ell}(\cos\theta)-\cos\theta
P'_{\ell}(\cos\theta)\right)\right) \\ \times \int d\omega \int d\omega' \frac{C
C'\omega\omega'}{16\pi}\cos\left((r+t+\delta)(\omega-\omega')\right)\,.
\label{eq:flux}
\end{multline}
Consider first the angular integral.  Let $x=\cos\theta$. The defining
differential equation for Legendre Polynomials is
\begin{equation} \label{LegendreEq}
 (1-x^2)P^{\prime\prime}_{\ell}(x) - 2xP'_{\ell}(x) + \ell(\ell+1)P_{\ell}(x) =
0\,.
\end{equation}
and the associated Legendre functions may be defined by
\begin{equation} \label{AssociatedEq}
 P^{(m)}_{\ell} = (-1)^m (1-x^2)^{m/2}\frac{d^m}{d x^m} P_{\ell}(x).
\end{equation}
 They satisfy the orthonormality conditions
\begin{equation} \label{LegendreNorm}
 \int_{-1}^1 dx P^{(m)}_k(x)P^{(m)}_{\ell} =
\frac{2(\ell+m)!}{(2\ell+1)(\ell-m)!}\delta_{k\ell}
\end{equation}
and
\begin{equation} \label{AssociatedNorm}
 \int_{-1}^1 dx \frac{P^{(n)}_{\ell}(x)P^{(m)}_{\ell}}{(1-x^2)} =
  \begin{cases}
     0, & m\neq n \\
   \tfrac{(\ell+m)!}{m(\ell-m)!}, & m = n \neq 0 \\
   \infty, & m=n=0
  \end{cases}.
\end{equation}
Using Eq.~(\ref{LegendreEq}) to rewrite $P^{\prime\prime}_{\ell}(x)$ in terms
of $P_{\ell}(x)$ and $P'_{\ell}(x)$, the angular integral in 
Eq.~(\ref{eq:flux}) becomes
\begin{equation} \label{AngularInt}
 \int_{-1}^1 dx \left[P^2_{\ell}(x)\left(1+\left(1-\ell(\ell+1)\right)^2\right)
- 2x^2 \left(P'_{\ell}(x)\right)^2 +
2\ell(\ell+1)xP'_{\ell}(x)P_{\ell}(x)\right].
\end{equation}
The first term satisfies the equal $\ell$ orthonormality condition, Eq.\
(\ref{LegendreNorm}), so
\begin{equation}
 \left(1+(1-\ell(\ell+1))\right)\int_{-1}^1 dx P^2_{\ell}(x) =
\left(1+(1-\ell(\ell+1))\right)\frac{2}{2\ell+1}.
\end{equation}
The second term is a little more complicated.  Begin by integrating by parts
\begin{equation}
 \int_{-1}^1 dx\, x^2 (P'_{\ell}(x))^2 = \left[x^2 P'_{\ell}(x)
P_{\ell}(x)\right|_{-1}^1  - \int_{-1}^1 dx \left[2x P_{\ell}(x)P'_{\ell}(x) +
x^2 P_{\ell}(x) P^{\prime\prime}_{\ell}(x)\right].
\end{equation}
Using Eq.\ (\ref{LegendreEq}), $x^2P^{\prime\prime}_{\ell}(x) =
P^{\prime\prime}_{\ell}(x)-2xP'_{\ell}(x)+\ell(\ell+1)P_{\ell}(x)$ and the
preceding is now
\begin{equation}
 \left[x^2 P'_{\ell}(x) P_{\ell}(x)\right|_{-1}^1 - \ell(\ell+1)\int_{-1}^1 dx
P_{\ell}(x)P_{\ell}(x) -\int_{-1}^1 dx P_{\ell}(x) P^{\prime\prime}_{\ell}(x).
\end{equation}
The second term here satisfies the orthonormality condition, while integrating
the third term by parts and using Eq.\ (\ref{AssociatedEq}) gives
\begin{equation}
 \left[x^2 P'_{\ell}(x) P_{\ell}(x)\right|_{-1}^1 - \left[P'_{\ell}(x)
P_{\ell}(x)\right|_{-1}^1 - \frac{2\ell(\ell+1)}{2\ell+1} + \int_{-1}^1 dx
\frac{P^{(1)}_{\ell}(x)P^{(1)}_{\ell}(x)}{(1-x^2)}\, ,
\end{equation}
where the
first and second terms cancel. Meanwhile, the last term satisfies 
Eq.~(\ref{AssociatedNorm}) and finally
\begin{equation}
 \int_{-1}^1 dx\, x^2 (P'_{\ell}(x))^2 = \frac{(\ell+1)!}{(\ell-1)!} -
\frac{2\ell(\ell+1)}{2\ell+1}.
\end{equation}
Consider now the last term of (\ref{AngularInt}).  An integration by parts gives
\begin{equation}
 \int_{-1}^1 dx\, xP'_{\ell}(x)P_{\ell}(x) =
\left[xP_{\ell}(x)P_{\ell}(x)\right|_{-1}^1 - \int_{-1}^1 dx\,
xP'_{\ell}(x)P_{\ell}(x) - \int_{-1}^1 dx P_{\ell}(x)P_{\ell}(x).
\end{equation}
The second term on the right is the same as the original integral, while the
third integral on the right satisfies Eq.\ (\ref{LegendreNorm}).  Furthermore,
$P_{\ell}(1) = 1$ while $P_{\ell}(-1) = (-1)^{\ell}$ so that
$\left[xP_{\ell}(x)P_{\ell}(x)\right|_{-1}^1 = 2$ and therefore
\begin{equation}
 \int_{-1}^1 dx\, xP'_{\ell}(x)P_{\ell}(x) = \frac{2\ell}{2\ell+1}.
\end{equation}
Adding up all the results gives
\begin{multline}
 \int_{-\pi}^{\pi} \sin\theta d\theta
\left(P^2_{\ell}(\cos\theta)\left(1+(1-\ell(\ell+1))^2\right) \right. \\ -
\left. 2\cos\theta P'_{\ell}(\cos\theta)\left(\sin^2\theta
P^{\prime\prime}_{\ell}(\cos\theta)-\cos\theta
P'_{\ell}(\cos\theta)\right)\right) \\ =
\frac{2}{2\ell+1}\left[2+\ell^2(\ell+1)(\ell+3)\right] -
\frac{2(\ell+1)!}{(\ell-1)!}
\end{multline}

The flux through the surface of the sphere is now
\begin{equation}
\Phi =  \varepsilon^{-1} \tilde{L} \int d\omega \int d\omega' \frac{C
C'\omega\omega'}{4}\cos\left((r+t+\delta)(\omega-\omega')\right)
\end{equation}
where 
\begin{equation}
 \tilde{L}=\frac{1}{2\ell+1}\left[2+\ell^2(\ell+1)(\ell+3)\right] -
\frac{(\ell+1)!}{(\ell-1)!}.
\end{equation}
The total energy of the wavepacket may now be found by integrating the flux
through the surface of the sphere for all time, $\int dt\Phi$.  Writing the
cosine function in terms of exponentials it is clear that $\int dt
\cos\left((r+t+\delta)(\omega-\omega')\right) =
2\pi\delta(\omega-\omega')\cos((r+\delta)(\omega-\omega'))$.  The delta function
takes care of the integration over $\omega'$, and setting the total energy to
$\omega/2$ ($\hbar=1$) gives the condition
\begin{equation}
 \varepsilon^{-1}\tilde{L}\int d\omega \frac{C^2\pi\omega^2}{2} =
\frac{1}{2}\omega.
\end{equation}
For a wavepacket sharply peaked in frequency, we may approximate
\begin{equation}
 \int_{j\varepsilon}^{(j+1)\varepsilon} \omega^2 d\omega \approx
\omega^2\Delta\omega = \omega^2\varepsilon,
\end{equation}
then
\begin{equation}
 A_{\ell m}(\omega) = C = \frac{1}{\sqrt{\pi\tilde{L}\omega}}.
\end{equation}

This normalization constant is the one contained in the radial function $Z(r)$,
so that when the normalized wavepacket is written
\begin{equation}
 \Psi_{(j)\mu\nu}^+ = \int_{j\varepsilon_j}^{(j+1)\varepsilon_j}
\frac{d\omega}{\sqrt{\pi\tilde{L}\varepsilon_j\omega}} e^{-i\omega\delta_j}
\Psi^+_{\mu\nu},
\end{equation}
it is understood that the radial function takes the value
$Z(r_*\to\infty)=e^{-i\omega r_*}$.


\section{Squeezed States} \label{app:SqueezedStates}
A squeezed
state is the natural state for a quantum mechanically created particle
occupying an in-vacuum state represented in an out-Fock space.
Squeezed quantum states are generated via the unitary displacement
and squeeze operators. This Appendix provides a brief summary of the
relevant ideas and results for squeezed states, primarily following
the notation found in Ref.~\cite{Caves:1981hw}; see also
Refs.~\cite{Stoler:1969tq,Stoler:1972sz,Yuen:1976vy}. 

Squeezed states are generate using the unitary displacement and squeeze
operators.  The displacement operator is
\begin{equation}
 D(\alpha)= e^{\alpha \hat{a}^{\dagger}-\alpha^*\hat{a}}.
\end{equation}
One can check that $D(\alpha)$ transforms $\hat{a}$ and $\hat{a}^{\dagger}$ as 
\begin{equation}
 D^{\dagger}(\alpha) \hat{a} D(\alpha) = \hat{a}+\alpha \quad \mathrm{and} \quad
D^{\dagger}(\alpha) \hat{a}^{\dagger} D(\alpha) = \hat{a}^{\dagger}+\alpha^*.
\end{equation}
The squeeze operator is
\begin{equation}
 \hat{S}(\zeta) =
\exp\left[\tfrac{1}{2}\zeta^*\hat{a}^2-\tfrac{1}{2}\zeta(\hat{a}^{\dagger}
)^2\right], \quad \zeta = \rho e^{i\theta}.
\end{equation}
Here the squeezing parameter, $\zeta$, is an arbitrary complex number. One may
show that the squeeze operator transforms $\hat a$ and $\hat a^{\dagger}$ as
\begin{subequations} \label{Sq:gp}
 \begin{equation} \label{Sq:gp1}
  S^{\dagger}(\zeta)\hat{a}S(\zeta) =
\hat{a}\cosh{\rho}-\hat{a}^{\dagger}e^{i\theta}\sinh{\rho} \\
 \end{equation}
and
 \begin{equation} \label{Sq:gp2}
    S^{\dagger}(\zeta)\hat{a}^{\dagger}S(\zeta) =
\hat{a}^{\dagger}\cosh{\rho}-\hat{a}e^{-i\theta}\sinh{\rho}.
 \end{equation}
\end{subequations}
\subsection{Multimode Squeezed State}
Suppose the state $|0,\zeta\rangle$ is a multimode squeezed state of
the form
\begin{equation}
 |0,\zeta\rangle = \prod_{\ell = n}^m S(\zeta_{\ell})|0\rangle.
\end{equation}
The expectation values under consideration require the calculation 
of terms such as $\sum_{j}\sum_{k}\langle\zeta,0\vert
\hat{a}_j\hat{a}_k\vert 0,\zeta\rangle$, $\sum_{j}\sum_{k}\langle\zeta,0\vert
\hat{a}_j\hat{a}_k\hat{a}_r\hat{a}_s\vert 0,\zeta\rangle$, etc.  
Consider for example
\begin{equation}
\langle\zeta,0\vert \hat{a}_j^{\dagger}\hat{a}_k\vert 0,\zeta\rangle = \langle
0|\prod_{\ell = n}^m S^{\dagger}(\zeta_{\ell}) \hat{a}_j^{\dagger}\hat{a}_k
\prod_{\ell = n}^m S(\zeta_{\ell})|0\rangle.
\end{equation}
This expands as
\begin{multline}
\langle 0|\left(
S^{\dagger}(\zeta_m)S^{\dagger}(\zeta_{m-1})...S^{\dagger}(\zeta_n)\hat{a}_j
S(\zeta_n)S(\zeta_{n+1})...S(\zeta_m)\right) \\ \times \left(
S^{\dagger}(\zeta_m)S^{\dagger}(\zeta_{m-1})...S^{\dagger}(\zeta_n)\hat{a}_k
S(\zeta_n)S(\zeta_{n+1})...S(\zeta_m)\right)|0\rangle.
\end{multline}
The action of $S(\zeta_{\ell})$ on $\hat{a}_j^{\dagger}$ and $\hat{a}_j$ is
\begin{subequations}
 \begin{equation}
  S^{\dagger}(\zeta_{\ell})\hat{a}_j^{\dagger}S(\zeta_{\ell}) =
(\hat{a}_j^{\dagger} \cosh\rho_{\ell} - \hat{a}_i\sinh\rho_{\ell})\delta_{j\ell}
+ \hat{a}_j^{\dagger}(1-\delta_{j\ell})
 \end{equation}
and
 \begin{equation}
  S^{\dagger}(\zeta_{\ell})\hat{a}_j S(\zeta_{\ell}) = (\hat{a}_j
\cosh\rho_{\ell} - \hat{a}_j^{\dagger}\sinh\rho_{\ell})\delta_{j\ell} +
\hat{a}_j(1-\delta_{j\ell}).
 \end{equation}
\end{subequations}
The action of a range of squeezing is then
\begin{equation}
 S^{\dagger}(\zeta_m)...S^{\dagger}(\zeta_n)\hat{a}_j^{\dagger}
S(\zeta_n)...S(\zeta_m) = (\hat{a}_j^{\dagger} \cosh\rho_{j} -
\hat{a}_j\sinh\rho_{j})\Theta_{nm}(j) + \hat{a}_j^{\dagger}(1-\Theta_{nm}(j)),
\end{equation}
where the integer step function is defined as
\begin{equation}
 \Theta_{nm}(j) = 
 \begin{cases}
 1, & n\leq j \leq m, \\ 
 0, & \text{otherwise}.
 \end{cases}
\end{equation}
It follows that
\begin{multline}
 \sum_{jk}\langle\zeta,0\vert \hat{a}_j^{\dagger}\hat{a}_k\vert 0,\zeta\rangle =
\sum_{jk} \langle\zeta,0\vert \left[(\hat{a}_j^{\dagger} \cosh\rho_{j} -
\hat{a}_j\sinh\rho_{j})\Theta_{nm}(j) +
\hat{a}_j^{\dagger}(1-\Theta_{nm}(j))\right] \\ \times \left[(\hat{a}_k
\cosh\rho_{k} - \hat{a}_k^{\dagger}\sinh\rho_{k})\Theta_{nm}(k) +
\hat{a}_k(1-\Theta_{nm}(k))\right] \vert 0,\zeta\rangle \\ =
\sum_{jk}\left[\left(-\Theta_{nm}(j)\sinh\rho_j\right)
\left(-\Theta_{nm}(k)\sinh\rho_k\right)\right]\delta_{jk}.
\end{multline}
To summarize, one finds
\begin{subequations} \label{doubleExp}
 \begin{equation} \label{Expaa}
  \sum_{jk}\langle\zeta ,0\vert \hat{a}_j\hat{a}_k\vert 0,\zeta\rangle =
\sum_{jk}\left[\left(1+\Theta_{nm}(j)(\cosh\rho_j-1)\right)\left(-\Theta_{nm}
\sinh\rho_k\right)\right]\delta_{jk},
 \end{equation}
 \begin{equation} \label{Expadagadag}
\sum_{jk}\langle\zeta,0\vert \hat{a}_j^{\dagger}\hat{a}_k^{\dagger}\vert
0,\zeta\rangle =
\sum_{jk}\left[\left(-\Theta_{nm}(j)\sinh\rho_j\right)\left(1+\Theta_{nm}
(k)\cosh\rho_k\right)\right]\delta_{jk},
 \end{equation}
 \begin{equation} \label{Expaadag}
  \sum_{jk}\langle\zeta ,0\vert \hat{a}_j\hat{a}_k^{\dagger}\vert 0,\zeta\rangle
= \sum_{jk} \left[\left(-\Theta_{nm}(j)\sinh\rho_j\right)
\left(-\Theta_{nm}\sinh\rho_k\right)\right]\delta_{jk},
 \end{equation}
and
 \begin{equation} \label{Expadaga}
  \sum_{jk}\langle\zeta ,0\vert \hat{a}_j^{\dagger}\hat{a}_k\vert 0,\zeta\rangle
= \sum_{jk} \left[\left(1+\Theta_{nm}(j)(\cosh\rho_j-1)\right)
\left(1+\Theta_{nm}(k)\cosh\rho_k\right)\right]\delta_{jk}.
 \end{equation}
\end{subequations}
Note that the result of Eq.\ (\ref{Expadaga}) contains a $\delta_{jk}$ that
makes the sum divergent.  Renormalization is therefore taken to correspond to
restricting the sum over modes to those occupying an excited squeezed state
mode, i.e.\ $\sum_{j=n}^m$.  For products of two operators this is equivalent to
normal ordering; equations (\ref{Expaa})-(\ref{Expadaga}) become
\begin{subequations}
  \begin{equation}
  \sum_{jk}\langle\zeta ,0\vert \hat{a}_j\hat{a}_k\vert 0,\zeta\rangle =
\sum_{jk}-\cosh\rho_j\sinh\rho_k\delta_{jk},
 \end{equation}
 \begin{equation}
\sum_{jk}\langle\zeta,0\vert \hat{a}_j^{\dagger}\hat{a}_k^{\dagger}\vert
0,\zeta\rangle = \sum_{jk}-\sinh\rho_j\cosh\rho_k\delta_{jk},
 \end{equation}
 \begin{equation}
  \sum_{jk}\langle\zeta ,0\vert \hat{a}_j\hat{a}_k^{\dagger}\vert 0,\zeta\rangle
= \sum_{jk} \sinh\rho_j\sinh\rho_k\delta_{jk},
 \end{equation}
and
 \begin{equation}
  \sum_{jk}\langle\zeta ,0\vert \hat{a}_j^{\dagger}\hat{a}_k\vert 0,\zeta\rangle
= \sum_{jk} \cosh\rho_j\cosh\rho_k\delta_{jk}.
 \end{equation}
\end{subequations}

The study of scalar field stress tensor induced fluctuations further requires
the use of the expectation value of four-operator products such as
\begin{subequations}
 \begin{multline}
 \sum_j\sum_k\sum_r\sum_s\langle\zeta,0\vert
\hat{a}_j\hat{a}_k\hat{a}^{\dagger}_r\hat{a}^{\dagger}_s\vert 0,\zeta\rangle =
\\ 
\sum_{jkrs}\langle 0\vert \left[(\hat{a}_j\cosh\rho_j - \hat{a}^{\dagger}_j
\sinh\rho_j )\Theta_{nm}(j) + \hat{a}_j (1-\Theta_{nm}(j))\right] \\ 
\times \left[(\hat{a}_k\cosh\rho_k - \hat{a}^{\dagger}_k \sinh\rho_k
)\Theta_{nm}(k) + \hat{a}_k (1-\Theta_{nm}(k))\right] \\ 
\times \left[(\hat{a}^{\dagger}_r\cosh\rho_r - \hat{a}_r \sinh\rho_r
)\Theta_{nm}(r) + \hat{a}^{\dagger}_r (1-\Theta_{nm}(r))\right] \\ 
\times \left[(\hat{a}^{\dagger}_s\cosh\rho_s - \hat{a}_s \sinh\rho_s
)\Theta_{nm}(s) + \hat{a}^{\dagger}_s (1-\Theta_{nm}(s))\right]\vert 0\rangle
\end{multline}
\begin{multline}
= \sum_{jkrs} \left[
(1+\Theta_{nm}(j)(\cosh\rho_j-1))(1+\Theta_{nm}(k)(\cosh\rho_k-1)) \right. \\
\times \left. (1+\Theta_{nm}(r)(\cosh\rho_r-1))
(1+\Theta_{nm}(s)(\cosh\rho_s-1))(\delta_{jr}\delta_{ks}+\delta_{js}\delta_{kr})
\right] \\ + \left[
(1+\Theta_{nm}(j)(\cosh\rho_j-1))(-\sinh\rho_k\Theta_{nm}(k)) \right. \\ \times
\left. (1+\Theta_{nm}(r)(\cosh\rho_r-1))
(-\sinh\rho_s\Theta_{nm}(s))\delta_{jk}\delta_{rs}\right].
\end{multline}
\end{subequations}
In this case there is the term
$((\delta_{jr}\delta_{ks}+\delta_{js}\delta_{kr})$ which makes the sum over
modes divergent.  Again, renormalization corresponds to restricting the sums
over modes to those modes which lie in the range of squeezing.  As discussed in
in Sect.~\ref{sec:ScalarField}, 
restricting the sum over modes this way for the four-operator products
corresponds, in the limit $\rho_i \gg 1$, 
to retaining only the fully normal ordered term.  The required
results are presented here, incorporating the restriction on the sum over modes,
but omitting the summation symbol for notational simplification.
\begin{subequations} \label{Eq:QuadExp}
 \begin{multline}
  \langle\zeta,0\vert \hat{a}_j\hat{a}_k\hat{a}_r\hat{a}_s \vert 0,\zeta\rangle
=
\cosh\rho_j\cosh\rho_k\sinh\rho_r\sinh\rho_s(\delta_{jr}\delta_{ks}+\delta_{js}
\delta_{kr})\\ +
\cosh\rho_j\sinh\rho_k\cosh\rho_r\sinh\rho_s(\delta_{jk}\delta_{rs}),
 \end{multline}
 \begin{multline}
  \langle\zeta,0\vert \hat{a}_j\hat{a}_k\hat{a}^{\dagger}_r\hat{a}^{\dagger}_s
\vert 0,\zeta\rangle =
\cosh\rho_j\cosh\rho_k\cosh\rho_r\cosh\rho_s(\delta_{jr}\delta_{ks}+\delta_{js}
\delta_{kr}) \\+
\cosh\rho_j\sinh\rho_k\cosh\rho_r\sinh\rho_s(\delta_{jk}\delta_{rs}),
 \end{multline}
 \begin{multline}
  \langle\zeta,0\vert \hat{a}_j\hat{a}_k\hat{a}_r\hat{a}^{\dagger}_s \vert
0,\zeta\rangle =
-\cosh\rho_j\cosh\rho_k\sinh\rho_r\cosh\rho_s(\delta_{jr}\delta_{ks}+\delta_{js}
\delta_{kr}) \\-
\cosh\rho_j\sinh\rho_k\cosh\rho_r\cosh\rho_s(\delta_{jk}\delta_{rs}),
 \end{multline}
 \begin{multline}
  \langle\zeta,0\vert \hat{a}_j\hat{a}_k\hat{a}^{\dagger}_r\hat{a}_s \vert
0,\zeta\rangle =
-\cosh\rho_j\cosh\rho_k\cosh\rho_r\sinh\rho_s(\delta_{jr}\delta_{ks}+\delta_{js}
\delta_{kr}) \\-
\cosh\rho_j\sinh\rho_k\sinh\rho_r\sinh\rho_s(\delta_{jk}\delta_{rs}),
 \end{multline}
 \begin{multline}
  \langle\zeta,0\vert \hat{a}^{\dagger}_j\hat{a}^{\dagger}_k\hat{a}_r\hat{a}_s
\vert 0,\zeta\rangle =
\sinh\rho_j\sinh\rho_k\sinh\rho_r\sinh\rho_s(\delta_{jr}\delta_{ks}+\delta_{js}
\delta_{kr}) \\+
\sinh\rho_j\cosh\rho_k\cosh\rho_r\sinh\rho_s(\delta_{jk}\delta_{rs}),
 \end{multline}
 \begin{multline}
  \langle\zeta,0\vert
\hat{a}^{\dagger}_j\hat{a}^{\dagger}_k\hat{a}^{\dagger}_r\hat{a}^{\dagger}_s
\vert 0,\zeta\rangle =
\sinh\rho_j\sinh\rho_k\cosh\rho_r\cosh\rho_s(\delta_{jr}\delta_{ks}+\delta_{js}
\delta_{kr}) \\+
\sinh\rho_j\cosh\rho_k\sinh\rho_r\cosh\rho_s(\delta_{jk}\delta_{rs}),
 \end{multline}
 \begin{multline}
  \langle\zeta,0\vert
\hat{a}^{\dagger}_j\hat{a}^{\dagger}_k\hat{a}_r\hat{a}^{\dagger}_s \vert
0,\zeta\rangle =
-\sinh\rho_j\sinh\rho_k\sinh\rho_r\cosh\rho_s(\delta_{jr}\delta_{ks}+\delta_{js}
\delta_{kr}) \\-
\sinh\rho_j\cosh\rho_k\cosh\rho_r\cosh\rho_s(\delta_{jk}\delta_{rs}),
 \end{multline}
 \begin{multline}
  \langle\zeta,0\vert
\hat{a}^{\dagger}_j\hat{a}^{\dagger}_k\hat{a}^{\dagger}_r\hat{a}_s \vert
0,\zeta\rangle =
-\sinh\rho_j\sinh\rho_k\cosh\rho_r\sinh\rho_s(\delta_{jr}\delta_{ks}+\delta_{js}
\delta_{kr}) \\-
\sinh\rho_j\cosh\rho_k\sinh\rho_r\sinh\rho_s(\delta_{jk}\delta_{rs}),
 \end{multline}
 \begin{multline} \label{Expaadagaa}
  \langle\zeta,0\vert \hat{a}_j\hat{a}^{\dagger}_k\hat{a}_r\hat{a}_s \vert
0,\zeta\rangle =
-\cosh\rho_j\sinh\rho_k\sinh\rho_r\sinh\rho_s(\delta_{jr}\delta_{ks}+\delta_{js}
\delta_{kr}) \\-
\cosh\rho_j\cosh\rho_k\cosh\rho_r\sinh\rho_s(\delta_{jk}\delta_{rs}),
 \end{multline}
 \begin{multline}
  \langle\zeta,0\vert
\hat{a}_j\hat{a}^{\dagger}_k\hat{a}^{\dagger}_r\hat{a}^{\dagger}_s \vert
0,\zeta\rangle =
-\cosh\rho_j\sinh\rho_k\cosh\rho_r\cosh\rho_s(\delta_{jr}\delta_{ks}+\delta_{js}
\delta_{kr}) \\-
\cosh\rho_j\cosh\rho_k\sinh\rho_r\cosh\rho_s(\delta_{jk}\delta_{rs}),
 \end{multline}
 \begin{multline}
  \langle\zeta,0\vert \hat{a}_j\hat{a}^{\dagger}_k\hat{a}_r\hat{a}^{\dagger}_s
\vert 0,\zeta\rangle =
\cosh\rho_j\sinh\rho_k\sinh\rho_r\cosh\rho_s(\delta_{jr}\delta_{ks}+\delta_{js}
\delta_{kr}) \\+
\cosh\rho_j\cosh\rho_k\cosh\rho_r\cosh\rho_s(\delta_{jk}\delta_{rs}),
 \end{multline}
 \begin{multline}
  \langle\zeta,0\vert \hat{a}_j\hat{a}^{\dagger}_k\hat{a}^{\dagger}_r\hat{a}_s
\vert 0,\zeta\rangle =
\cosh\rho_j\sinh\rho_k\cosh\rho_r\sinh\rho_s(\delta_{jr}\delta_{ks}+\delta_{js}
\delta_{kr}) \\+
\cosh\rho_j\cosh\rho_k\sinh\rho_r\sinh\rho_s(\delta_{jk}\delta_{rs}),
 \end{multline}
 \begin{multline} \label{Expadagaaa}
  \langle\zeta,0\vert \hat{a}^{\dagger}_j\hat{a}_k\hat{a}_r\hat{a}_s \vert
0,\zeta\rangle =
-\sinh\rho_j\cosh\rho_k\sinh\rho_r\sinh\rho_s(\delta_{jr}\delta_{ks}+\delta_{js}
\delta_{kr}) \\-
\sinh\rho_j\sinh\rho_k\cosh\rho_r\sinh\rho_s(\delta_{jk}\delta_{rs}),
 \end{multline}
 \begin{multline}
  \langle\zeta,0\vert
\hat{a}^{\dagger}_j\hat{a}_k\hat{a}^{\dagger}_r\hat{a}^{\dagger}_s \vert
0,\zeta\rangle =
-\sinh\rho_j\cosh\rho_k\cosh\rho_r\cosh\rho_s(\delta_{jr}\delta_{ks}+\delta_{js}
\delta_{kr}) \\-
\sinh\rho_j\sinh\rho_k\sinh\rho_r\cosh\rho_s(\delta_{jk}\delta_{rs}),
 \end{multline}
 \begin{multline}
  \langle\zeta,0\vert \hat{a}^{\dagger}_j\hat{a}_k\hat{a}_r\hat{a}^{\dagger}_s
\vert 0,\zeta\rangle =
\sinh\rho_j\cosh\rho_k\sinh\rho_r\cosh\rho_s(\delta_{jr}\delta_{ks}+\delta_{js}
\delta_{kr}) \\+
\sinh\rho_j\sinh\rho_k\cosh\rho_r\cosh\rho_s(\delta_{jk}\delta_{rs}),
 \end{multline}
and
 \begin{multline}
  \langle\zeta,0\vert \hat{a}^{\dagger}_j\hat{a}_k\hat{a}^{\dagger}_r\hat{a}_s
\vert 0,\zeta\rangle =
\sinh\rho_j\cosh\rho_k\cosh\rho_r\sinh\rho_s(\delta_{jr}\delta_{ks}+\delta_{js}
\delta_{kr}) \\+
\sinh\rho_jk\sinh\rho_k\sinh\rho_r\sinh\rho_s(\delta_{jk}\delta_{rs}).
 \end{multline}
\end{subequations}

\bibliography{enhanced}

\end{document}